\documentclass[twocolumn, floats,revtex4-1,nofootinbib,usenatbib]{emulateapj}
\usepackage{graphicx}
\usepackage{bm}
\usepackage{natbib}
\usepackage{hyperref}
\usepackage[english]{babel}
\usepackage{tabu}
\usepackage{amsmath}
\usepackage{float}
\usepackage{color}
\usepackage{hyperref}

\usepackage[flushleft]{threeparttable}

\pdfoutput=1 

\begin{document}

\title{Calibrating star-formation rate prescriptions at different scales (10 pc to 1 kpc) in M31}

\author{Neven Tomi\v{c}i\'{c}$^{1}$}
\author{I-Ting Ho$^{1}$}
\author{Kathryn Kreckel$^{1}$}
\author{Eva Schinnerer$^{1}$}
\author{Adam Leroy$^{2}$}
\author{Brent Groves$^{3}$}
\author{Karin Sandstrom$^{4}$}
\author{Guillermo A. Blanc$^{5}$}
\author{Thomas Jarrett$^{6}$}
\author{David Thilker$^{7}$}
\author{Maria Kapala$^{6}$}
\author{Rebecca McElroy$^{1}$}

\affil{
$^1$Max Planck Institute for Astronomy (MPIA), K\"onigstuhl 17, 69117 Heidelberg, Germany
$^2$Department of Astronomy, The Ohio State University, 4055 McPherson Laboratory, 140 West 18th Avenue, Columbus, Ohio, USA
$^3$Australian National University, Canberra ACT 2600, Australia
$^4$Center for Astrophysics and Space Sciences, Department of Physics, University of California, San Diego, 9500 Gilman Drive, La Jolla, CA 92093, USA
$^5$Observatories of the Carnegie Institution for Science, 813 Santa Barbara Street, Pasadena, California, 91101, USA
$^6$University of Cape Town, Private Bag X3, Rondebosch 7701, South Africa
$^7$Center for Astrophysical Sciences, The Johns Hopkins University,  Zanvyl Krieger School of Arts \& Sciences, 3400 N. Charles St, Baltimore,  21218 Maryland, USA
}

\email{tomicic@mpia-hd.mpg.de}
\date{\today}

\shorttitle{Calibrating the SFR prescription in M31}
\shortauthors{Tomi\v{c}i\'{c} et al.}

\begin{abstract}

We calibrate commonly used star formation rate (SFR) prescriptions using observations in five kpc-sized fields in the nearby galaxy Andromeda (M31) at 10\,pc spatial resolution. Our observations at different scales enable us to resolve the star-forming regions and to distinguish them from non star-forming components. We use extinction corrected H$\alpha$  from  optical integral field spectroscopy as our reference tracer and have verified its reliability via tests. It is used to calibrate monochromatic and hybrid (H$\alpha$+a$\times$IR and FUV+b$\times$IR) SFR prescriptions, which use FUV (GALEX), 22\,$\mu$m (WISE) and 24\,$\mu$m (MIPS). Additionally, we evaluate other multi-wavelength infra-red tracers.  Our results indicate that the SFR prescriptions do not change (in M31) with spatial scales or with subtraction of the diffuse component. For the calibration factors in the hybrid SFR prescriptions, we find a$\approx$0.2 and  b$\approx$22 in M31, which are a factor of 5 higher than in the literature.  As the fields in M31 exhibit high attenuation and low dust temperatures, lie at large galacto-centric distances, and suffer from high galactic inclination compared to measurements in other galaxies, we propose that the fields probe a dust layer extended along the line of sight that is not directly spatially associated with star-forming regions. This (vertically) extended dust component increases the attenuation and alters the SFR prescriptions in M31 compared to literature measurements. We recommend that SFR prescriptions should be applied with caution at large galacto-centric distances and in highly inclined galaxies, due to variations in the relative (vertical) distribution of dust and  gas.

\end{abstract}


\section{Introduction}
\label{sec:intro}

 Star formation (SF) affects the chemical evolution and distribution of stars and the interstellar medium (ISM), and thus the morphology and evolution of galaxies (\citealt{Schmidt59}, \citealt{Kennicutt97}, \citealt{Kennicutt03}, \citealt{Murphy11}). 
 Therefore, understanding the rate and location of star formation  is crucial. Reliable star formation rate (SFR) tracers are needed to properly quantify the SF activity. 
 
 Various SFR tracers  target direct or reprocessed light from short-lived massive, young and luminous stars (e.g. see \citealt{Kennicutt07}). 
 Some of the  widely adopted SFR tracers of the luminous stars and its surrounding ionized gas are: ultraviolet (UV) stellar continuum emission, nebular hydrogen emission lines (e.g.~Balmer H$ \alpha $) and free-free radio continuum emission. 
 However, the stellar and nebular light can be obscured by the dust, and usually underestimate the true SFR (\citealt{Thilker07,Calzetti07,Kennicutt07}). The light absorbed by dust is re-radiated as infrared (IR) emission (\citealt{Gao04}, \citealt{Calzetti05}, \citealt{Calzetti07}, \citealt{Kennicutt07}, \citealt{Rieke09}, \citealt{Murphy11},  \citealt{HCRodriges15}). 
 The mid-IR tracers alone also underestimate SFR. Therefore, combining the obscured (UV and optical) and un-obscured (IR) tracers results in the total SFR (\citealt{Kennicutt03,Calzetti05, Wu05, Calzetti07,Thilker07,Tabatabaei10,Leroy12, Davis14,CatalanTorrecilla15}). 
 The combination of tracers are often referred to as "hybrid" SFR prescriptions, while single tracers are "monochromatic" prescriptions.
  
Much of the work done in determining the calibration of monochromatic or hybrid SFR prescriptions, which use emission lines from the ionized gas,  has two major caveats.
The first caveat is that the imaging of emission lines based on broad-band and narrow-band filters does not account for underlying stellar absorption lines, or for contamination from neighboring emission lines. 
However, recent progress in integral field unit (IFU; \citealt{Barden88}) spectroscopy resolves individual spectral lines and the underlying stellar continuum, enabling accurate mapping of Balmer emission lines  (\citealt{Kapala15, CatalanTorrecilla15,Davies16}). 

The second caveat comes from the low spatial resolution of existing studies, which mostly probe the ISM at 0.5-1 kpc scales or at galactic scales (\citealt{Kennicutt97},\citealt{Calzetti05},\citealt{Salim07},\citealt{Thilker07},\citealt{Jarrett12},\citealt{Leroy12},\citealt{CatalanTorrecilla15},\citealt{Jarrett17}). 
For comparison, active star-forming regions (i.e.~HII regions) typically have sizes between 30\,pc and  200\,pc (\citealt{Issa81,Azimlu11}). 
Hence, extragalactic studies of the SFR rate have been unable to resolve HII regions, mixing HII regions and regions without star formation, e.g. diffuse ionized gas (DIG; \citealt{Haffner09}) and "IR-cirrus"\footnote{IR-cirrus refers to the diffuse component in mid-IR images, which corresponds to emission re-radiated by dust heated by older stellar populations (\citealt{Leroy12}).}.
A contribution from the DIG and mid-IR cirrus, and additional ultra-violet emission from older stars may change the SFR prescriptions. Moreover, different regions within the kpc-sized beam may differ in their physical (temperatures, stellar and ISM densities), chemical (metallicities) and morphological (distribution) conditions.
For example, \citet{Eufrasio14} observed the interacting galaxy NGC 6872 with 10 kpc size apertures and found variations in the FUV-IR conversion factor that are correlated with regional differences in stellar populations.
Similarly, \citet{Boquien16} studied eight nearby galaxies at kpc scales and found that the SFR prescription changes with stellar surface density, rather than with the dust attenuation or SFRs. 
The best spatial resolution (at ~30\,pc) achieved by extragalactic studies of SFR are from observing galaxies in the local group (\citealt{Hony15, BoquienCalzetti15}).

In this work, we will calibrate the SFR prescription by using maps of different SFR tracers (FUV, H$\alpha$, and IR) in 5 fields of the Andromeda galaxy (M31).
Due to its proximity ($\approx0.78$\,Mpc), we can achieve good spatial resolution (10 pc). 
Our fields are  0.6\,kpc$\times$0.9\,kpc in projected size, which enables us to test the  SFR prescriptions at various spatial scales and to spatially resolve the HII regions.
In addition, we will use IFU spectral data in order to map multiple Balmer lines and measure the extinction corrected H$\alpha$ emission to use  as our baseline SFR tracer.

There are a few main goals of this paper. 
First, we will test the reliability of extinction corrected H$\alpha$ as a SFR tracer. 
Secondly, we will study the behavior of different monochromatic and hybrid SFR tracers (22\,$\rm \mu $m, 24\,$\rm \mu $m, H$ \alpha $+22 $\rm \mu $m, H$ \alpha $+24 $\rm \mu $m, FUV+22 $\rm \mu $m, FUV+24 $\rm \mu $m, 12\,$\rm \mu $m, 70\,$\rm \mu $m, 160\,$\rm \mu $m and total infrared) at different spatial scales (from 10\,pc to 750\,pc). 
We will also test how  the diffuse components affect SFR prescriptions. Finally, we will test if dust temperature and (3-dimensional) dust/gas distributions play a role in changing the SFR prescriptions in M31. 

The paper is structured as follows. 
In section \ref{Sec:Data}, we describe how the data are calibrated and present maps of different star formation tracers. In section \ref{Sec:Ha}, we test the reliability of the extinction corrected H$\alpha$ (labeled as H$\alpha,\rm corr$) as a SFR tracer. 
We present the main results of our comparisons and provide prescriptions for SFR from the monochromatic and the hybrid tracers in Section \ref{Sec:SFR Calibrations}. 
In the same section, we also test the effects of varying spatial scales and subtracting the diffuse emission. 
In section \ref{Sec: aIR vs InclR25}, we demonstrate a possible connection between dust temperature and the SFR prescriptions. 
Discussion and summary are in Sections \ref{Sec:Discussion} and \ref{Sec:Summary}.

\begin{figure}[t!]
\centering
\includegraphics[width=1.0\linewidth]{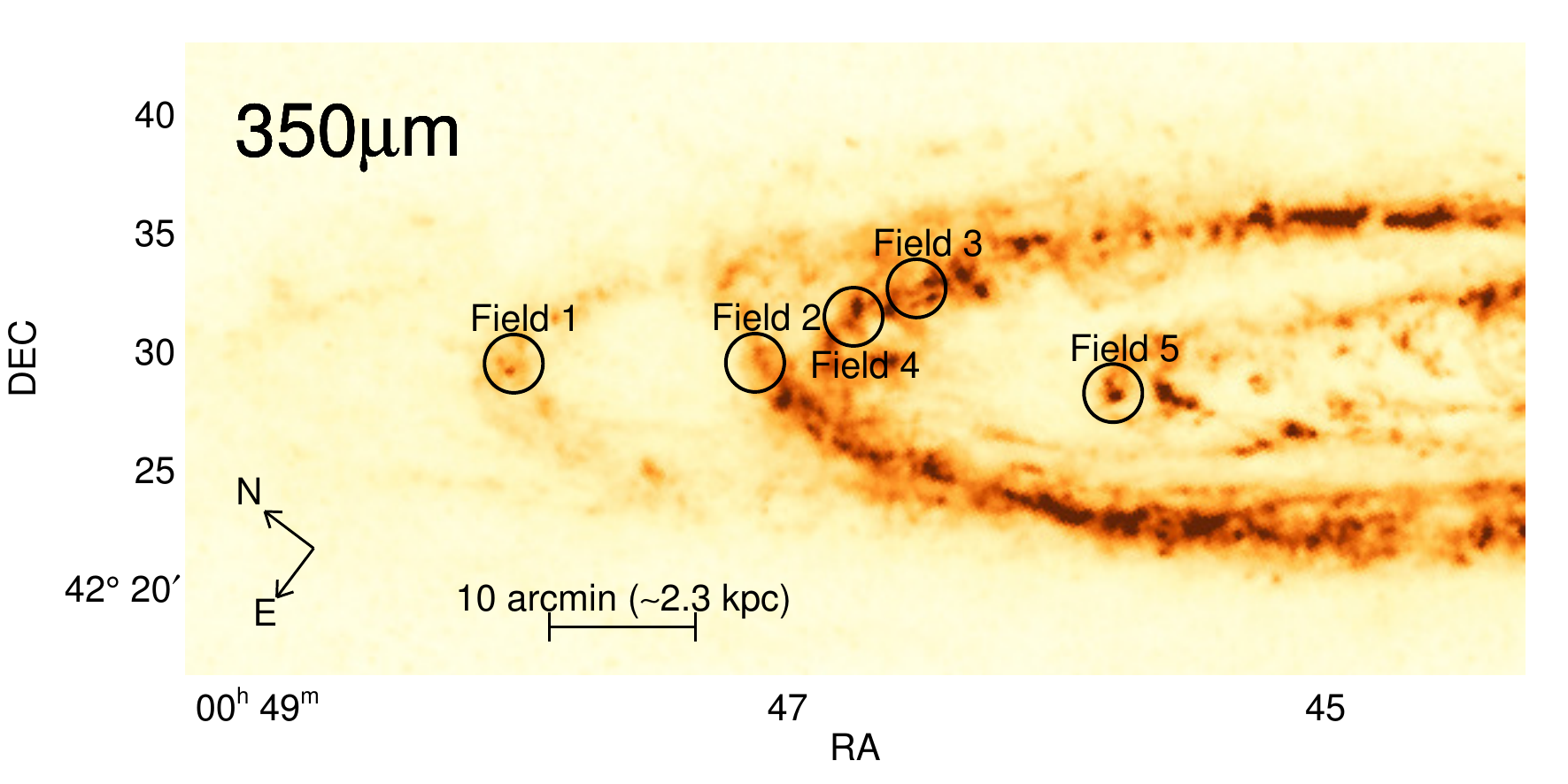}
\caption{ SPIRE 350 $\rm \mu $m intensity map of M31 and positions of the five fields used in this work. The map has 24.9'' resolution (corresponding to $\approx100$\,pc at the distance of M31).}
\label{fig:FigM31}
\end{figure}

 \begin{table}[t!]
\centering
\caption{Coordinates, approximate distances from the galaxy center (in kpc) and metallicities (using equation 5 in \citealt{Zurita12}) for our fields.  }
\begin{tabular}{ccccc}
\hline 
    	     Field & R.A. & Dec. & R & Z \\ 
    	     &  (J2000) &  (J2000) & kpc & 12+log(O/H)\\ 
    	   \hline 1 &$ 00^{h}46^{m}28.88^{s}$ & $+42^{\circ}11'38.16''$ & 16 & 8.3  \\ 
    	          2 &$ 00^{h}45^{m}34.04^{s}$ & $+41^{\circ}58'33.53''$ & 12.2 & 8.4 \\ 
    	          3 &$ 00^{h}44^{m}36.04^{s}$ & $+41^{\circ}52'53.58''$ & 11.7 & 8.4 \\ 
    	          4 &$ 00^{h}44^{m}58.54^{s}$ & $+41^{\circ}55'09.14''$ & 11.8  & 8.4 \\ 
    	          5 &$ 00^{h}44^{m}25.58^{s}$ & $+41^{\circ}37'37.20''$ & 6.8 & 8.6  \\ 
    	   \hline 
    	   \\
\end{tabular}  \\
\label{tab:Tab01}
\end{table}

\section{Data}\label{Sec:Data}

  M31 is a nearby ($ \sim $780\,kpc; \citealt{Stanek98})  and massive (stellar mass of $\approx10^{11}$\,M$_{\odot}$; \citealt{Geehan06}) SA(s)b galaxy (\citealt{Corwin94}), which makes observing the interstellar medium (ISM) at high spatial resolution possible. The inclination of the galaxy is $\sim$77$^\circ$ (\citealt{Henderson79}, \citealt{Courteau11} , \citealt{Dalcanton12}) and R$ _{25}\approx $20.5\,kpc \footnote{R$ _{25}$ is the radius at which the observed optical intensity is equal to 25 mag in the B band.}  \citep{Zurita12}. The galaxy also shows ring-like structures at galactocentric radii of 6, 10 and 15 kpc (\citealt{Gordon06}).

  We use integral field unit (IFU) data from 5 fields (each with a projected size of $\approx$600\,pc $ \times $ 900\,pc; see Fig. \ref{fig:FigM31} and \ref{fig:SFTmaps_F1}), chosen to cover a range of star formation rates and environments. 
  Positions, radial distances, and metallicities of the 5 fields are tabulated in Table \ref{tab:Tab01} and shown in Figure \ref{fig:FigM31}.  

  The IFU spectroscopic data provide H$ \alpha $ line maps that are combined with 22\,$\rm \mu $m, 24\,$\rm \mu $m and FUV images. 
  We use FUV emission for our SFR calibration instead of NUV because FUV traces  younger stars ($ < $30 Myr old), while NUV can also be emitted by older stars.  We adopt WISE maps for the 22\,$\rm \mu $m \citep{Wright10}, MIPS maps for the 24 $\rm \mu $m data (Spitzer; \citealt{Rieke04}, \citealt{Engelbracht07}), and GALEX (\citealt{Martin05}) for the FUV data.
  A benefit of using the GALEX and WISE observations for FUV and 22\,$\rm \mu $m  is that thay are from all-sky surveys so the derived calibrations can be applied, taking into account the caveats discussed in this paper, to other extragalactic  objects in the sky.
  Additionally, we will also calibrate other tracers, including 12\,$\rm \mu $m (near-IR tracer of PAH\footnote{Polycyclic aromatic hydrocarbon molecules.}), 70\,$\rm \mu $m, 160\,$\rm \mu $m (tracing cold dust) and total IR (TIR).  
  All images are Nyquist-sampled with $\approx$3 pixels across the instrumental point spread function (PSF). 
  We refer to the full width at half maximum (FWHM) of the PSF as the native angular resolution for each tracer. 
  The observed wavelength,  best achieved angular resolution, and pixel size for each instrument are listed in Table \ref{tab:Tab02}. 
  The units of the FUV and mid-IR images are flux densities per pixel (F$ _{\lambda} $ for FUV and F$ _{\nu} $ for IR). 
   The final intensity maps (in units of erg\,s$ ^{-1} $\,cm$ ^{-2} $\,arcsec$ ^{-2} $), used in this work, are defined as F$ _{\lambda}  $    (F$ _{\nu} $) maps that are devided by their pixel sizes (in arcsec$ ^2 $) and multiplied by their effective wavelengths (frequencies).
 For the calibration of the SFR prescriptions, we will consistently use and show de-projected surface brightness values of the tracers throughout this work, assuming M31's inclination of $\sim$77$^\circ$ \footnote{We multiply the area of each pixel or aperture by a factor of 4 to correct the minor axis for the inclination. We estimate corrections  following Eq. 1 in \citet{Bergh88}. } .

\begin{table*}[t!]
\begin{threeparttable}

\centering
\caption{Instrument, wavelength coverage, effective wavelength, angular resolution, spatial resolution and size of pixel of datasets used.    }

\begin{tabular}{cccccc}
\hline 
Instrument & $\Delta\lambda $ & $ \lambda_{eff}$ &  angular res. & spatial res. & original size of pixel \\ 
Name &  ($\rm \mu $m) & ($\rm \mu $m) & (arcsec) & (pc) & (arcsec/pix) \\ 
\hline 
PPaK IFU (Calar Alto)\tnote{a} &$ 0.37-0.7$ &$ 0.5 $ &2.7 & 10  &1 \\ 
W4, 22\,$\rm \mu $m (WISE)\tnote{b} &$20-26$  &$22.1 $ &$ 11.9 $ & 45  &4.4 \\ 
MIPS 24\,$\rm \mu $m (Spitzer)\tnote{c} &$21.5-26.2$  &$ 24 $ &$ 6.4 $ & 24  &2.4 \\ 
FUV (GALEX)\tnote{d} &$0.135-0.175$  &$ 0.154 $ &$ 4.5 $ & 17  & 1.5\\ 
 \hline 
& & Additional tracers: & & &\\
W3, 12\,$\rm \mu $m (WISE)\tnote{b} &$7-18$  &$11.6 $ &$ 6.6 $ & 23  &2.2 \\   
PACS 70\,$\rm \mu $m \tnote{e} &$60-85$  &$71 $ &$ 5.7 $ & 20  &2 \\   
PACS 160\,$\rm \mu $m \tnote{e} &$120-210$  &$160 $ &$ 11 $ & 40  &4.2 \\   
  \hline 
    	   \\ 
\end{tabular}  
 \begin{tablenotes}
 \footnotesize
\item[a] SLIM survey; \citet{Kapala15}.
\item[b] Jarrett et al. 2018; submitted.
\item[c] \citet{Gordon06}.
\item[d] \citet{Thilker05}.
\item[e] \citet{Groves12}.
\end{tablenotes}
\end{threeparttable}

\label{tab:Tab02}
\end{table*}

 \subsection{WISE 22 $\rm \mu $m and SPITZER 24 $\rm \mu $m images}
 \label{Subsec: IR images}
 
For the 22 $\rm \mu $m images,  we use maps from the Wide Field Infrared Survey Explorer (WISE; \citealt{Wright10}). The six-degree-wide maps were constructed to preserve the native resolution of WISE  W4 images  using a drizzle technique (\citealt{Jarrett12}).  As described in \citet{Chauke14} and   \citet{Jarrett17},  foreground Galactic stars were identified and removed, and the satellite galaxies M32 and  M110 were  subtracted from the final set of images.   The mean background "sky" level was measured 2.8$ ^\circ $ radius from the center of M31, and globally subtracted from the final set of images. For the flux calibration, we used the prescription given by \citet{Cutri11}, while using 7.871~Jy as the flux value for Vega (\citealt{Brown14, Jarrett17}). The uncertainty maps are composed and calculated from instrumental flat-field errors (1\% of intensity value), Poisson errors and the sky background errors. 

The 24\,$\rm \mu $m images are from the MIPS instrument on the {\em Spitzer} Space Telescope (\citealt{Rieke04}, \citealt{Werner04},\citealt{Engelbracht07}). We use the maps presented by \citealt{Gordon06}. Unlike the 22 $\rm \mu $m maps, the PSF of the  24\,$\rm \mu $m maps presents bright secondary Airy  rings (\citealt{Rieke04}, \citealt{Kennicutt07}, \citealt{Engelbracht07}, \citealt{Temim10}, \citealt{Aniano11}). These may present a problem when analyzing ISM features on the 24 $\rm \mu $m maps at the highest resolutions. After carefully analyzing the shape of the PSF, we conclude that 90\% of the flux of the source is contained within the first Airy ring.  

 We check how similar the 22\,$\rm \mu $m and 24\,$\rm \mu $m maps are, to  evaulate whether the hybrid prescriptions would change when using different mid-IR tracers.  The comparison shown in Fig. \ref{fig:22vs24} demonstrates a tight correlation, implying that the two mid-IR maps match when convolved to the same resolutions. That is expected because the instruments' filters have a similar wavelength coverage (\citealt{Wright10, Jarrett11}). However, we find that the 22\,$\rm \mu $m  data has 0.03 dex higher flux densities compared to 24\,$\rm \mu $m, with 0.05 dex scatter. A small fraction ($\approx5\%$ or less) of the pixels  are brighter ($\approx0.1$\,dex brighter) in 24$ \mu $m than in 22$ \mu $m.  This minor difference could be due to the different PSFs of the two instruments. We conclude that the hybrid SFR prescription would not change appreciably if we replace one mid-IR tracer with the other. 
 
 In the following work, we will convolve the maps to larger spatial resolutions and use integrated intensities in apertures with a minimum radius of 13'' to better sample the entire flux of compact sources (\citealt{Kennicutt07}).

  \subsection{Other IR tracers: WISE 12$\,\mu$m, PACS 70$\,\mu$m and PACS 160$\,\mu$m data}
  \label{Subsec: 12 and PACS images}

In this work, we will test the SFR prescription as a function of the dust temperature and of the fraction of emission from the cold dust, which may indicate if our data originate from the dust  dominantly heated by old stellar population instead of HII regions (\citealt{Groves12,Ford13}). PACS 70$\,\mu$m and PACS 160$\,\mu$m maps (\citealt{Poglitsch10}) are used to measure the 70$\,\mu$m/160$\,\mu$m ratio, which traces the dust temperature,  and the 160$\,\mu$m/TIR ratio, which indicates the fraction of  emission from the cold dust. The reduction procedure is described in \citet{Groves12}. The noise level of the PACS 70$\,\mu$m (160$\,\mu$m) maps is $7.5\cdot10^{-5}$\,Jy\,arcsec$^{-2}$ ($1.6\cdot10^{-3}$Jy\,arcsec$^{-2}$). We subtract the background using 10 apertures (R$\approx$90'') outside the galaxy (at least $\approx5$\, arcmin from the second ring in M31).  We also use the WISE 12\,$\rm \mu $m map (W3 band), which is calibrated in a similar way as the WISE 22 $\rm \mu $m map (described in Sec. \ref{Subsec: IR images}). 
 
\subsection{GALEX FUV data}
\label{Subsec: UV images}

The FUV mosaic images were observed with GALEX (The Galaxy Evolution Explorer; \citealt{Martin05}). Details of the observations and calibration are described in \citet{Thilker05}, \citet{Morrissey07} and \citet{Thilker07}.  

For the sky subtraction, we use 100 apertures (75''$ \times $75'' in size) positioned around M31 (minimum of 5 arcmin from the second ring in M31). The mode of 100 aperture mean values is used as the background value, which is  $\approx1.3\cdot10^{-15}\,$erg\,s$^{-1}$\,cm$^{-2}$\,arcsec$^{-2}$ for the FUV images. The noise level of the FUV images is  $\approx2\cdot10^{-16}\,$erg\,s$^{-1}$\,cm$^{-2}$\,arcsec$^{-2}$. Additionally, we correct the UV maps for MW foreground extinction using the \citet{Cardelli89} extinction curve (CCM), assuming R$ _{\rm V}  = $3.1 (\citealt{Clayton15}) and using E$ _{\rm B-V} =$0.055 (\citealt{Schlafly11}). \citet{Peek13} found that the foreground extinction of FUV should be 30\% higher compared to the extinction derived by the CCM extinction curve. If we apply that correction, the extinction corrected FUV emission in M31 would increase by $\approx$10\% (0.05 dex), and only have a minor effect on calibrating the SFR prescriptions.  For the uncertainty maps, we follow the prescription described by \citet{Morrissey07} and \citet{Thilker05}. The background sky uncertainty ($\approx2\cdot10^{-17}\,$erg\,s$^{-1}$\,cm$^{-2}$\,arcsec$^{-2}$) is added in quadrature to the instrumental uncertainty.

\begin{figure*}[t!]
\centering
\includegraphics[width=1.\linewidth]{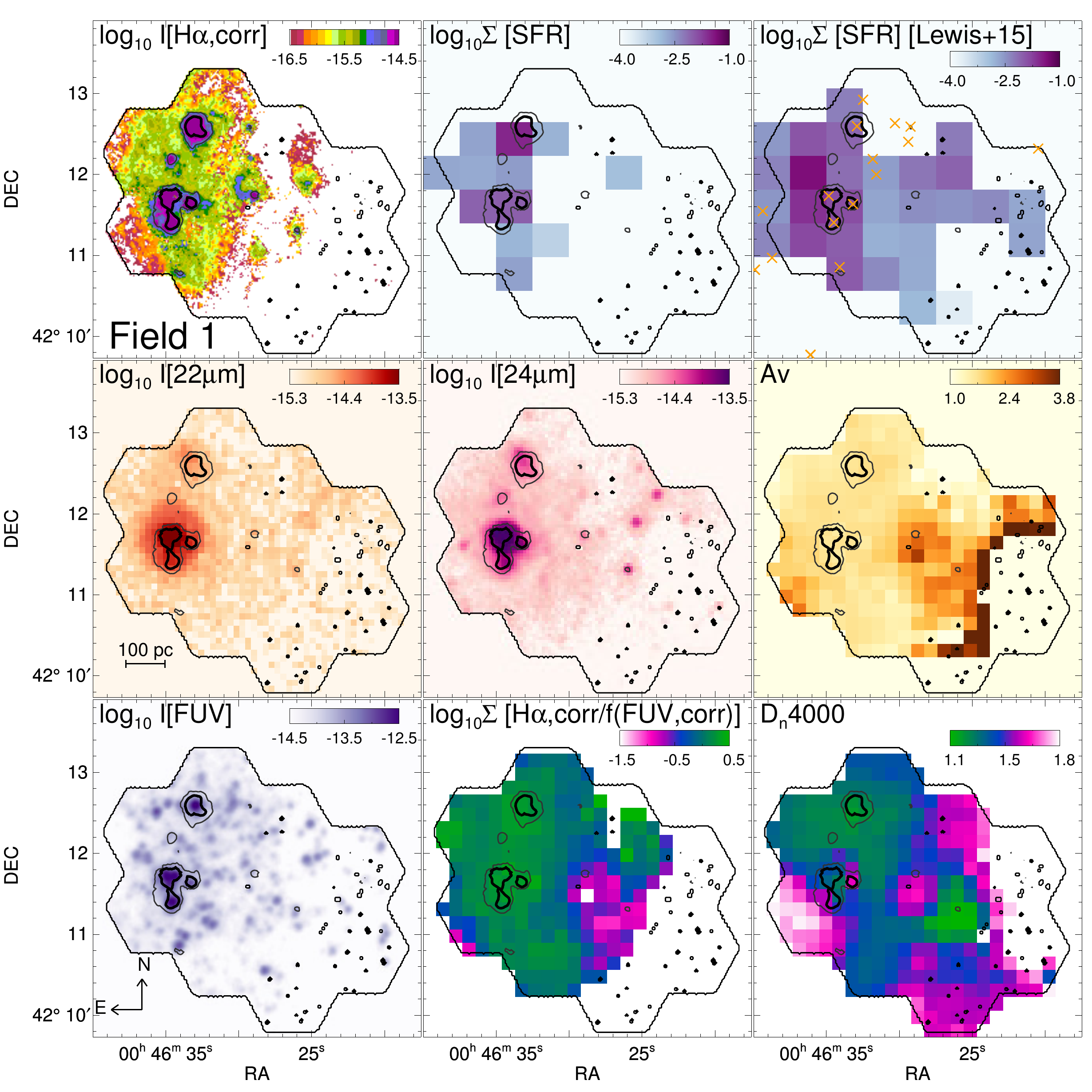}
\caption{ Maps of Field 1 showing: $\rm \Sigma$(H$ \alpha,corr $) (at native resolution; top row left), $\rm \Sigma_{SFR}(H\alpha,corr)$ from our spectra (pixel sizes of 23'' or 100\,pc; top row middle), $\rm \Sigma_{SFR}$ from the modeled star formation history by \citet{Lewis15} (pixel sizes of 23'' or 100\,pc; top row right), $\rm \Sigma$(22\,$ \mu $m) (at native resolution; middle row left), $\rm \Sigma$(24\,$ \mu $m) (at native resolution; middle row middle), A$\rm _V$ (at Spire 360\,$\rm \mu$m resolution; middle row right), $\rm \Sigma$(FUV\,$\rm \mu $m) (at native resolution; bottom row left), the  H$\alpha$,corr/f$\rm _\nu(FUV,corr)$ ratio (at Spire 360\,$\rm \mu$m resolution; bottom row middle), and the D$\rm _n4000$ break (at Spire 360\,$\mu$m resolution and estimated from the spectra; bottom row right). Contours on all images correspond to observed H$\alpha$ intensities of $ 3\times10^{-16} $ (thin) and  $ 10^{-15}  $(thick) erg\,s$ ^{-1} $cm$ ^{-2}$\,arcsec$^{-2}$ at native resolution. Discussion about the maps can be found in Sec. \ref{Subsec:SFTMaps}, and a comparison between  $\rm \Sigma_{SFR}(H\alpha,corr)$ and $\rm \Sigma_{SFR}$ from \citet{Lewis15} in Sec. \ref{Subsec:Lewis}. We added positions of young ($<30$\,Myr) stellar clusters identified by \citet{Fouesneau14} and \citet{Johnson16} as yellow crosses in upper right panel. }
\label{fig:SFTmaps_F1}
\end{figure*}

 \begin{figure}[t!]
\centering
\includegraphics[width=0.9\linewidth]{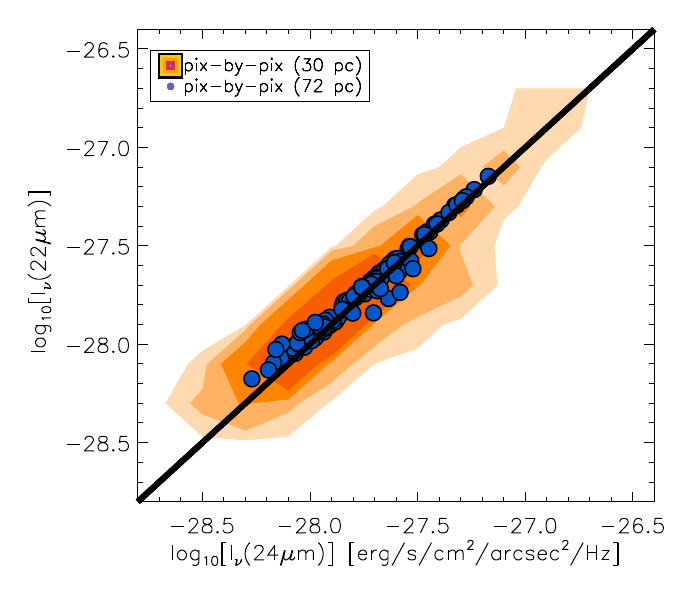}
\caption{  Pixel-by-pixel comparison between the intensities of MIPS 24$ \mu $m (x-axis)  and  WISE 22$ \mu $m (y-axis), at 25'' (red contours) and 65'' (blue filled circles) resolution. The 1-to-1 line is plotted in black. Generally, the data from the two instruments are equal except at lower fluxes. Some data points show slightly higher 24 $ \mu $m  values. The difference between the two images decreases with increasing pixel size.For details about the IR maps, see Sec. \ref{Subsec: IR images}.  }
\label{fig:22vs24}
\end{figure}

\subsection{Optical H$ \alpha $ data}
\label{Sec: Ha calib}

The optical IFU spectral data in this work were previously used by \citet{Kapala15}, \citet{Kapala17} and \citet{Tomicic17} as a part of the Survey of Lines in M31 (SLIM) project. 
The observation and calibration of the  data and the derivation of Balmer emission lines are described in detail in \citet{Kreckel13, Kapala15, Tomicic17}. Here we only provide a short summary.

 The observations were conducted using  the Potsdam Multi-Aperture Spectrophotometer (PMAS, \citealt{Roth05}) and specialized fiber-bundle PPaK  mode with the V300 grating (\citealt{Kelz06}) on the 3.5\,m telescope at the Calar Alto Observatory in 2011. 
 For each of the five fields ($ 3'\times4' $ or 680\,pc $ \times $ 900\,pc in size) a mosaic of 10 pointings with three dither positions were observed (overall 50 pointings for the entire galaxy).  
 Sky pointings away from the galaxy were observed  for sky subtraction, and  twilight sky fields are used for flat fielding. 
 Additionally, spectra of  the calibration continuum lamp and He+HgCd arc lamps were obtained for wavelength calibration, and  standard stars were observed for flux calibration (\citealt{Oke90}).
 We reduced and calibrated the data using the P3D software package\footnote{http://p3d.sourceforge.net/}, version 2.2.6. \citep{Sandin10} that applies the standard calibration techniques for IFU data. 
 The final reduced 2D spectra  are resampled onto a grid of 1\,arcsec$^2$ pixels, referred to henceforth as the data cubes. The data cubes have spectral resolution of R$=$1000 (centered in 5400 \AA{}), a wavelength range of 3700-7010 \AA{} and an angular resolution of 2.7$''$.
 Errors from the data and sky contribution are propagated through the entire calibration process.

 To obtain the fluxes of strong nebular emission lines, we analyze the reduced spectra using GANDALF\footnote{\textbf{G}as \textbf{And} \textbf{A}bsorption \textbf{L}ine \textbf{F}itting; http://www-astro.physics.ox.ac.uk/~mxc/software/} (version 1.5; \citealt{Sarzi06}). GANDALF fits both the stellar continuum  and the nebular emission lines iteratively using  penalized pixel-fitting (pPXF; \citealt{Cappellari04}). 
 The stellar template spectra are taken from the \citet{Bruzual03} simple stellar population (SSP) templates (\citealt{Tremonti04}), which span a range of stellar ages (5 Myr to 12 Gyr) and metallicities ($ Z=0.004 $ and  $ 0.05  $). 
 We assume the same $E_{\rm B-V}$ for the foreground extinction as for the FUV maps (Sec. \ref{Subsec: UV images}). 
 Foreground stars are also removed during the fitting.  In the following maps and diagrams,  we exclude pixels where the line fluxes do not pass the threshold of AoN\footnote{(AoN) values correspond to the ratio of the line amplitude and the noise.
 Noise for AoN is calculated as a standard deviation of residuals from the observed and the fitted spectra.} $>$ 3 (\citealt{Sarzi06}). 
 
 Our astrometry is computed and checked by comparing compact HII regions with those from the Local Group of Galaxies Survey\footnote{Maps are publicly available on  http://www2.lowell.edu/users/massey/lgsurvey.html and https://ned.ipac.caltech.edu/} (\citealt{Massey07,Azimlu11}), and by comparing \textit{r} and \textit{g} band SDSS images with our reduced maps.
 We find a maximum deviation of 1$ '' $. 
 Our flux calibration is checked by comparing SDSS \textit{r} band images to bandpass matched images extracted from the IFU spectra. 
 We estimate that our flux calibration is accurate within 0.06 dex scatter for the bright regions ($ >7\times10^{-18} $\, erg\,s$^{-1}$\,cm$^{-2}$\,arcsec$^{-2}$) and is offset 0.11 dex with 0.08 dex scatter for the low-brightness regions ($ <7\times10^{-18} $\, erg\,s$^{-1}$\,cm$^{-2}$\,arcsec$^{-2}$). 
Additionally, we compared narrow-band \textit{H$\alpha$+NII} images from the Survey of the Local Group of Galaxies to the H$\alpha$+NII bandpass matched images extracted from the IFU spectra. 
While comparing these images, we assumed NII/H$\alpha$=0.4 ratio (as in \citealt{Azimlu11}), although our spectral analysis indicates that ratios actually range from 0.2 to 0.6. 
The narrow-band images agree within 0.1 dex scatter in the bright regions ($ >5\times10^{-16} $\,erg\,$ ^{-1} $\,cm$ ^{-2} $\,arcsec$ ^{-2} $), and are offset by 0.07 dex with 0.15 dex scatter for  the  low-brightness regions ($ <5\times10^{-16} $\,erg\,$ ^{-1} $\,cm$ ^{-2} $\,arcsec$ ^{-2} $). 
These offsets are consistent with the quoted uncertainties from the literature (\citealt{Massey07,Azimlu11}).

\subsection{Convolution \& apertures} 
\label{Sec: Convolution}

In this paper, we will show the impact of varying spatial scales on the SFR tracers and the  SFR prescriptions. The calibration of the SFR prescriptions is done using two approaches. 

 The first compares pixels in the maps at matched angular resolution. We also test SFR prescriptions using integrated fields, i.e.~treat an entire field (with a projected size of $\rm \approx0.6\,kpc\times0.9\,kpc$) as one single aperture.  
 When we change the resolution of the maps, we convolve and re-bin the maps using convolution kernels, pipeline, and procedures from \citet{Aniano11}. In the case of IFU data cubes, we convolve and re-bin the optical maps in each wavelength channel before applying spectral analysis on the resulting convolved data cubes.  The integrated fields data from Fields 2 and 5 are not used in this work for calibrating the SFR prescriptions  due to their relatively low surface brightness and correspondingly low signal to noise ratio of H$\beta$. 
 
 The second approach uses apertures with matched radii, applied to the maps at their native resolutions. 
 We choose the positions of the apertures by eye, targeting regions with bright peaks in the SFR tracer maps  and a few regions dominated by diffuse emission. 
 The purpose of the apertures is to distinguish between star-forming and non-star-forming regions, and to be able to extract the diffuse emission outside all apertures.  
 We test the SFR prescriptions with apertures that have   radii of 13.5'', 27'' and 55'' (corresponding to $\approx$\,50\,pc, $\approx$\,100\,pc, $\approx$\,200\,pc in physical scales, respectively). 
For apertures placed on the optical data cubes, we convolve the cubes to the native resolution of the IR instrument, integrate spaxels at each wavelength channel within the aperture, and then apply spectral analysis on the resulting spectra. 
The PSF resolutions, pixel sizes, and aperture radii used for these measurements are tabulated in Table \ref{tab:Tab03}. The aperture positions are shown in Fig. \ref{fig:Apertures_fields}.

 \begin{table}[t!]
\centering
\caption{Top: Resolutions and projected spatial sizes of the maps used. Bottom: Radii of apertures used.  }
\begin{tabular}{ccc}
\hline{}
 & Maps & \\
 Resolution (arcsec) & pixel size (pc) &  pixel size (arcsec/pix) \\ 
 \hline 
 $ \approx $\,6 & 7 & 2.4 \\
 11.7 & 13 & 4 \\
 25 & 30 & 10 \\
 65 & 72 & 20\\
 Integrated & 600-900  & 180-270\\
\hline{}
 &Apertures & \\ 
\end{tabular}  \\
 \centering
 \begin{tabular}{cc}
 Radius in arcsec  &  projected  radius in pc \\
 \hline 
 13.5 & 50 \\
 27 & 100 \\
 55 & 200 \\
  \\
\end{tabular}  \\
\label{tab:Tab03}
\end{table}

\subsection{Maps of SFR tracers}  
\label{Subsec:SFTMaps}

Fig. \ref{fig:SFTmaps_F1} shows  Field 1 in all the star formation tracers (H$\alpha$,corr, FUV, 22\,$\rm \mu $m and 24\,$\rm \mu $m).  
Tracer maps for the other four fields are shown in Sec. \ref{Sec:Appendix_A_Maps} (Fig. \ref{fig:SFTmaps_F2},  \ref{fig:SFTmaps_F3},  \ref{fig:SFTmaps_F4}, and  \ref{fig:SFTmaps_F5}). 
Additionally, in those figures we show observed $\rm \Sigma_{SFR}(H\alpha)$, modeled $\rm \Sigma_{SFR}$ from  \citet{Lewis15} (see Sec. \ref{Subsec:Lewis}), A$\rm _V$, H$\alpha$,corr/f$\rm _\nu(FUV,corr)$ ratio maps, and the D$\rm _n4000$ break (estimated from the spectra and using the wavelength range as in Tab. 1 in \citealt{Balogh99}).
The H$\alpha$,corr/f$\rm _\nu(FUV,corr)$ ratio and the  D$\rm _n4000$ break are independent probes of the stellar age. 
The H$\alpha$,corr/f$\rm _\nu(FUV,corr)$ ratio \citealt{Leitherer99,Whitmore11,SanchezGill11}) decreases with higher age of the clusters.  
However, a direct conversion between the H$\alpha$,corr/f$\rm _\nu(FUV,corr)$ ratio and the age is highly uncertain and dependent on assumptions of initial mass functions (IMFs), metallicities, and spatial scales. 
Similar to the H$\alpha$,corr/f$\rm _\nu(FUV,corr)$ ratio, the  D$\rm _n4000$ break indicates a luminosity weighted age of stars, with higher values indicating older stellar populations. 
The D$\rm _n4000$ break is defined by \citet{Bruzual83} as a ratio of the fluxes in the stellar continuum at longer and shorter wavelengths from $\lambda4000$\,\AA.

 Most of the bright HII regions, visible in the H$\alpha$ maps, correspond to young stellar clusters  with their emission dominated by O and B stars that ionize their surrounding gas. 
The  maps show a good spatial correlation between the H$\alpha$  emission,  the H$\alpha$,corr/f$\rm _\nu(FUV,corr)$ ratio and the D$\rm _n4000$ break.
 We confirm that young stellar clusters lie in the centers of the bright HII regions using the PHAT catalog (\citealt{Dalcanton12}) of young   ($<30$\,Myr) clusters  from \citet{Fouesneau14} and  \citet{Johnson16}. 
 HII regions show a discrete and clumpy distribution throughout the fields. 
 Between and around the HII regions, we observe  diffuse H$\alpha$ emission that corresponds to the diffuse ionized gas (DIG, \citealt{Walterbos94}, \citealt{Haffner09}, \citealt{Tomicic17}), and it can be seen up to 200\,pc away from the HII regions. 
The mid-IR tracers show similar diffuse features. 
 
 Bright HII regions are well correlated with FUV and mid-IR emission. 
 However, mid-IR and FUV maps reveal additional low-brightness features that are not correlated with H$\alpha$ emission.
Moreover, some regions have bright H$\alpha$ emission and low intensity mid-IR emission, such as the bright northern  HII region in Field 1.  The FUV maps reveal a clumpy distribution around HII regions.
 Those FUV clumps do not show NUV emission, which excludes the possibility that it comes from less massive MW  foreground stars.
 While mid-IR maps show a relatively smooth distribution, there are some mid-IR regions that are not seen on the H$\alpha$ map. 
 These spatial variations between different tracers could indicate different stages in the time evolution of the clusters (\citealt{Whitmore11,SanchezGill11}).
 For example, the presence of mid-IR emission without H$\alpha$ may indicate a single embedded cluster within highly attenuated HII regions, while the reverse could be due to more evolved HII regions around OB associations. Lastly, FUV regions without mid-IR or H$\alpha$ emission could point to evolved old stellar populations that do not ionize the gas or heat the dust around them.

\section{H$ \alpha $ as our reference SFR tracer}\label{Sec:Ha}

 One advantage of using IFU spectra is that we can separate the nebular emission lines from the underlying stellar continuum with proper estimation of the underlying absorption. Additionally, we can map the attenuation of the H$\alpha$ line using the Balmer decrement. This combination allows us to spatially map the SFR at high physical resolution in M31, using  extinction corrected H$\alpha$ (H$\alpha,\rm corr$) as our reference SFR tracer. In this work, we use this measure, $\Sigma_{\rm SFR}$(H$\alpha,\rm corr$), as our fiducial SFR surface density.

\begin{figure*}[t!]
\centering
\includegraphics[width=0.8\linewidth]{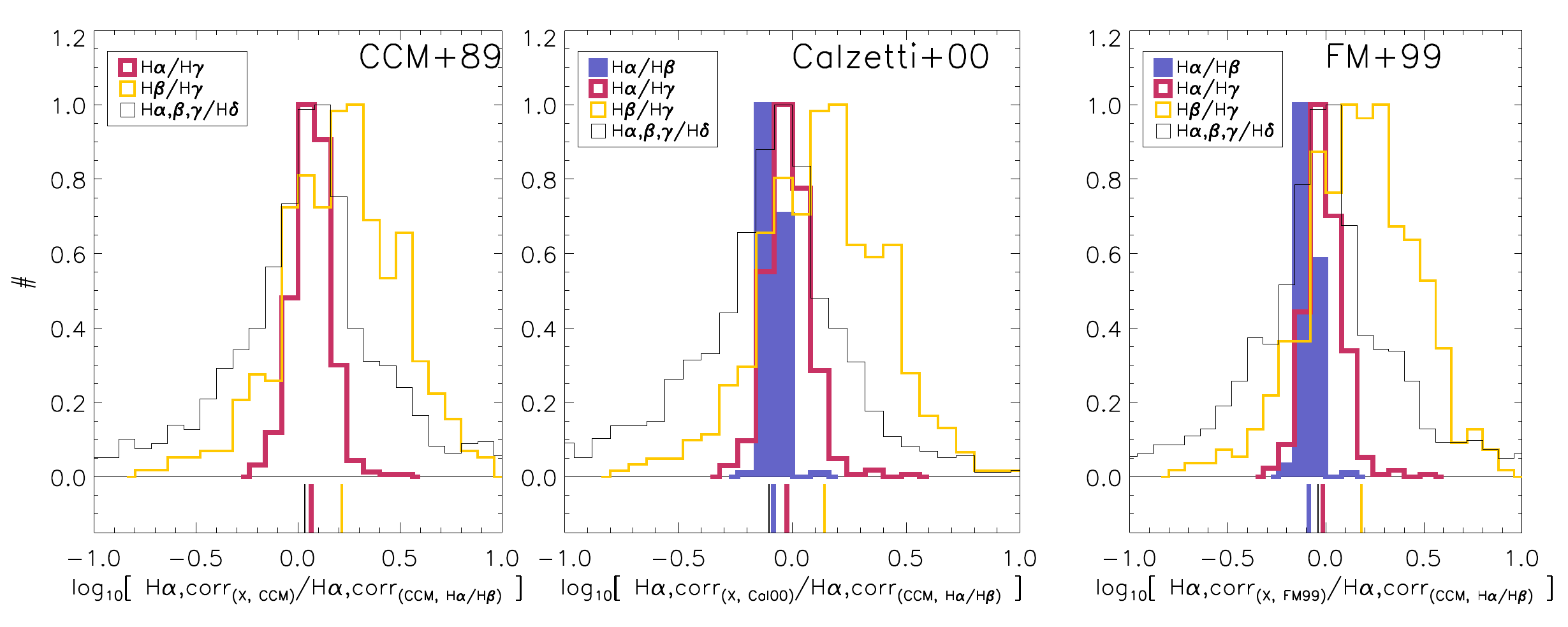}
\caption{ Histograms of ratios between $ \Sigma(\rm H\alpha,corr) $ derived  from different extinction curves and different line ratios, and our reference $ \Sigma(\rm H\alpha,corr) $ that uses the \citet{Cardelli89} extinction curve (CCM), R$_V$=3.1, and the H$\alpha$/H$\beta$ ratio. The purpose of this diagram is to see how much SFRs  based on H$\alpha,\rm corr$ deviate when we use different extinction curves or different Balmer line ratios. The extinction curves used within each panel are from: CCM (left panel), \citet{Calzetti00} (middle panel) and \citet{Fitzpatrick09} (labeled as FM+99; right panel). We assume R$_V$=3.1 for all curves. Different histograms utilize different line ratios: H$\alpha$/H$\beta$ (purple filled histogram), H$\alpha$/H$\gamma$ (thick red line), H$\beta$/H$\gamma$ (thin yellow line) and the line ratios with H$\delta$ (thin black line). Median values of the corresponding distributions are presented as vertical lines below the histograms. If we use the \citet{Fitzpatrick09} or \citet{Calzetti00} curve with H$\alpha$/H$\beta$, H$\alpha,\rm corr$ would deviate by 0.1 dex and would have a small uncertainty. The $ \Sigma(\rm H\alpha,corr) $ data derived from the other Balmer line ratios have more scatter due to the larger uncertainties in the line ratios and lines themselves. However, using  H$\alpha$/H$\gamma$ instead of H$\alpha$/H$\beta$ still gives an uncertainty of only $ \approx0.15 $dex.  All data are from  pixels with AoN$ > $5 for all Balmer lines.  }
\label{fig:Test_Av_histo}
\end{figure*}

 \begin{figure}[t!]
 \centering
\includegraphics[width=0.8\linewidth]{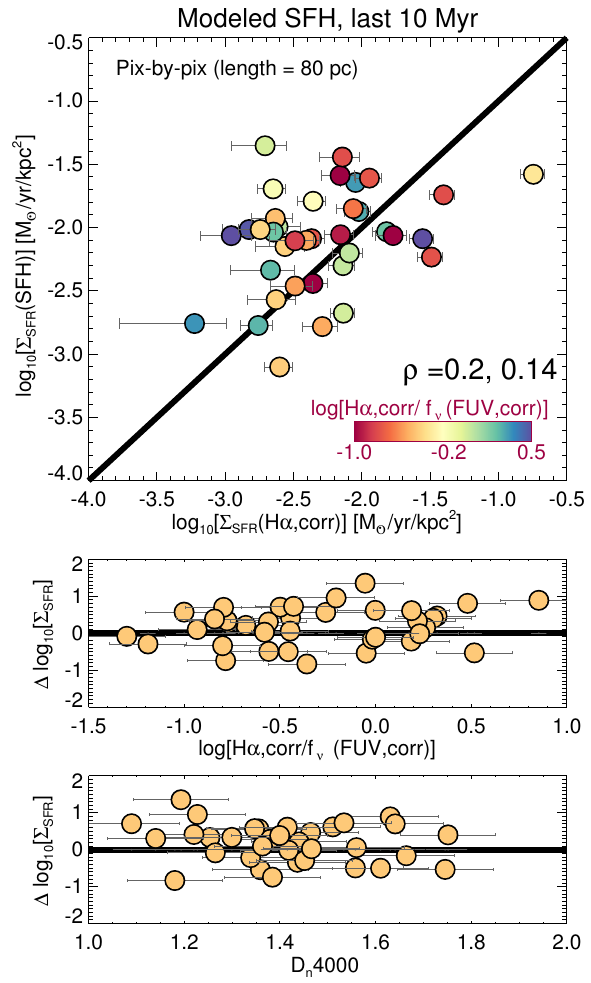}
\caption{ Comparison between our reference  $ \Sigma_{SFR}(\rm H\alpha,corr) $ from spectral fitting and the $ \Sigma_{SFR}(\rm SFH)$ in M31 derived from the modeled star formation history (SFH) averaged over the last 10 Myr by \citet{Lewis15}. The pixel-by-pixel  data points correspond to pixel sizes of 23'' ($\approx 80$\,pc).  The 1-to-1 relation lines are plotted as solid black lines, and we label the Spearman's  correlation coefficient ($ \rho $, left number) and  the significance of its deviation from zero (right number). The data are color-coded by the H$\alpha$,corr/f$\rm _\nu(FUV,corr)$ ratio. In the lower panels we show deviation of the data from the 1-to-1 relation  as a function of  the H$\alpha$,corr/f$\rm _\nu(FUV,corr)$ ratio and the luminosity weighted D$\rm _n4000$ break. Lower H$\alpha$,corr/f$\rm _\nu(FUV,corr)$ ratios and higher D$\rm _n4000$ break values indicate older stellar clusters (or populations).    }
\label{fig:Lewis}
\end{figure}

\subsection{Conversion from H$ \alpha $ and FUV to SFR}
\label{Subsec:HaToSFR}

H$\alpha,\rm corr$ emisison serves as a proper estimate of the SFR if two major criteria are fulfilled. The first criterion is that the extinction corrected H$\alpha,\rm corr$ flux recovers all intrinsic H$\alpha$ emission. The second important criterion is that the theoretical prescription for SFR estimation from H$\alpha,\rm corr$ flux is valid. The conversion from H$\alpha,\rm corr$ flux to SFR is well established under certain assumptions and widely used in the literature (for example in \citealt{Kennicutt98}, \citealt{Kennicutt03}, \citealt{Calzetti05}, \citealt{Leroy12}, \citealt{Murphy11}). It is derived under the assumptions that all ionizing radiation is absorbed, and that $ \approx $45\% of the ionized hydrogen atoms emit H$ \alpha $ photons during recombination (case B). It also assumes that the gas is purely ionized by young massive stars, and the stellar IMF is fully sampled. The duration of the star formation  should also be taken into account. A constant star formation rate will lead to  different H$ \alpha $/FUV ratios and different mid-IR emission behavior compared to the case of a single aged starburst. In previous papers, the continuous star formation assumption held because of sampling large spatial scales (often the entire galactic disks)  that encompasses multiple star forming regions (\citealt{Murphy11}). However, that assumption may be incorrect when observing smaller spatial scales (\citealt{Faesi14,Koepferl17}). 

 In this paper, we will adopt the H$ \alpha,\rm corr $-to-SFR conversion from \citet{Murphy11} that uses the Starburst99\footnote{http://www.stsci.edu/science/starburst99/docs/default.htm} stellar population models:
\begin{equation}
\frac{\rm \Sigma_{SFR}(H\alpha,\rm corr)}{\rm M_{\odot}yr^{-1}kpc^{-2} }=5.37\times10^{-42}\frac{\rm \Sigma(H\alpha,\rm corr)}{\rm erg\,s^{-1}kpc^{-2}}
\label{eq:Eq01}
\end{equation}
This conversion assumes a constant star formation over 100 Myr, a Kroupa initial mass function (IMF; \citealt{Kroupa01}), solar metallicities, Case B recombination and a gas temperature of $\approx10^{4}$\,K (for details see \citealt{Murphy11}). If we assume that mostly the young, massive and short lived stars ($<$20 Myr)  contribute significantly to the ionizing flux, then we can assume that this conversion factor is relatively independent of the previous star formation history and different time scales of star formation (\citealt{Kennicutt98,Murphy11}).   

 To derive $ \Sigma_{SFR}(FUV,corr) $ from $\Sigma(\rm FUV,corr)$, we use the following prescription from \citet{Murphy11}: 
\begin{equation}
\frac{\rm \Sigma_{SFR}(FUV,corr)}{\rm M_{\odot}yr^{-1}kpc^{-2} }=4.42\times10^{-44}\frac{\rm \Sigma(FUV,corr)}{\rm erg\,s^{-1}kpc^{-2}}
\label{eq:Eq02}
\end{equation}
One caveat of this method is that the older stellar population may contribute to the FUV emission, and that this conversion is variable with different time scales (\citealt{Kennicutt98,Murphy11}). The assumed timescale of star formation for this prescription is 100 Myr. However, this prescription may differ given the small spatial scales probed in M31 (\citealt{Faesi14}).

 In our previous paper, \citet{Tomicic17}, we found that the dust/gas distribution in M31 mostly follows the foreground screen models, and that the dust scale height is larger than the scale height of the DIG and HII regions for our studied fields. Therefore, in the following calculation, we will also assume a simple screen model of the dust/gas distribution, use the CCM extinction curve, the Balmer decrement of H$\alpha$/H$\beta$=2.86, and the selective extinction with R$_V =$3.1 (\citealt{Kreckel13}, \citealt{Tomicic17}).

\subsection{Effects of different extinction curves}
\label{Subsec:Av}

We test and show in  Fig.  \ref{fig:Test_Av_histo} the deviation in $ \Sigma(\rm H\alpha,corr) $  when using different extinction curves. 
All histograms  represent a comparison between $ \Sigma(\rm H\alpha,corr) $ derived from different extinction curves (CCM, \citealt{Calzetti00}, \citealt{Fitzpatrick09}) and our reference $ \Sigma(\rm H\alpha,corr) $ calculated from the CCM extinction curve, R$_V$=3.1, and using the H$\alpha$/H$\beta$ ratio. 
In all cases we assume R$_V$=3.1, due to the similarity in extinction curves and R$_V$ observed  between the MW and M31 by \citet{Clayton15}. 
While different panels show   different extinction curves, each individual histogram shows results using the ratio of different Balmer lines  (H$\alpha$/H$\beta$, H$\alpha$/H$\gamma$, H$\beta$/H$\gamma$, and the ratio of all Balmer lines and H$\delta$ line). 
For this test, we use pixels where AoN$ > $5 for all considered Balmer lines. 

For all extinction curves tested, $ \Sigma(\rm H\alpha,corr) $ calculated from H$ \alpha $/H$ \beta $ shows the smallest scatter and smallest offset from our reference $ \Sigma(\rm H\alpha,corr) $  that is estimated using the CCM extinction curve. 
This is   due to the higher S/N of the H$ \alpha $ and H$ \beta $ lines. 
The offset is only  $ 0.1 $ dex from the reference $ \Sigma(\rm H\alpha,corr) $, with a  scatter of $<$0.1 dex. 
Higher scatter is seen in the histograms that use the line ratios with  weaker Balmer lines (ratios with H$ \gamma $ and H$ \delta $ lines). 
This is due to larger uncertainties and the small wavelength difference of those lines with H$ \beta $, leading to higher systematic deviations in the line ratios and attenuation values. 
However, using  H$\alpha$/H$\gamma$ instead of H$\alpha$/H$\beta$ ratios still gives an  uncertainty of only $ \approx0.15 $ dex.  

 Our conclusion from these histograms is that the H$ \alpha $/H$ \beta $ ratio is more reliable than other line ratios, and that using different extinction curves in M31 with this ratio would change derived $ \Sigma_{SFR}(\rm H\alpha,corr) $ values by a maximum of $ 0.1 $ dex.

\subsection{Comparison with SFRs derived from the PHAT survey}
\label{Subsec:Lewis}

 We compare our $ \Sigma_{SFR}(\rm H\alpha,corr) $ values with those derived independently from resolved stellar photometry in M31. 
 \citet{Lewis15} and \citet{Lewis17} modeled spatially resolved star formation histories (SFH), A$_V$, and extinction corrected FUV emission (from integrated SFH)  using \textit{Hubble Space Telescope} (HST) images of M31 from the PHAT survey (\citealt{Johnson12}). 
 Their $\Sigma_{SFR}$(SFH)  maps are derived by integrating the modeled SFH over the last 10 Myr in each pixel of their M31 map. Note that the most recent time bin available in their model is 4 Myr. 
 To compensate for the lack of SFH on timescales shorter than 4 Myr, they estimated it by extrapolating from the  time bin  between  5 Myr and 6 Myr. 

Examples of our $ \Sigma_{SFR}(\rm H\alpha,corr) $ and the $ \Sigma_{SFR}$(SFH) maps are shown in Fig. \ref{fig:SFTmaps_F1}, \ref{fig:SFTmaps_F2}, \ref{fig:SFTmaps_F3}, \ref{fig:SFTmaps_F4}, and \ref{fig:SFTmaps_F5}. 
 While there is good agreement overall, the limitations of such a comparison to $\Sigma_{SFR}$(SFH) is visible just north-east of the bright southern HII region in Field 1, where there is a peak in $\Sigma_{SFR}$(SFH) that is offset from both the H$\alpha$ and mid-IR peaks. 
The FUV emission at the location of the peak in $\Sigma_{SFR}$(SFH) strongly suggests that here $\Sigma_{SFR}$(SFH) traces SF older than 5 Myr. 
Furthermore, areas  outside the HII regions and with old stellar clusters  (evident from the H$\alpha$,corr/f$\rm _\nu(FUV,corr)$ ratio and the spectral fit) have estimated $\Sigma_{SFR}$(SFH), while lacking H$\alpha$ emission.
We could not estimate $ \Sigma_{SFR}(\rm H\alpha,corr) $ for those regions. 
These spatial variations highlight the different sensitivity of both SFR tracers to the age of the star-forming regions, thus we restrict our comparison on Fig. \ref{fig:Lewis} to regions where information from both methods is available.

Fig. \ref{fig:Lewis} shows a  pixel-by-pixel comparison of the maps, where we re-bin our $\rm \Sigma_{SFR}(\rm H\alpha,corr) $ maps to spatially match the   $ \Sigma_{SFR} $(SFH) map, with pixel size of 23'' ($\approx70$\,pc). 
The data are color-coded by the  H$\alpha$,corr/f$\rm _\nu(FUV,corr)$ ratio, with older clusters having lower values. 
Although, $ \Sigma_{SFR} $(SFH) exhibits slightly higher ($\approx0.2$ dex) values than $\rm \Sigma_{SFR}(\rm H\alpha,corr) $,  there is a large scatter (standard deviation of 0.5 dex and a variation of up to $\approx$1 dex). 
We do not find any correlation of the residuals with age.  
This comparison is robust, but the large scatter in the data could be due to: a)   high uncertainties in the modeling of the recent SFH, b) uncertainty in the interpolation and estimation of the SFH in the last 4 Myr, and c) a gradual drop in H$\alpha$ emission on time scales longer than 5 Myr.

\subsection{Comparison by using molecular clouds masses}
\label{Subsec:Viaene}

\begin{figure}[t!]
\centering
\includegraphics[width=0.9\linewidth]{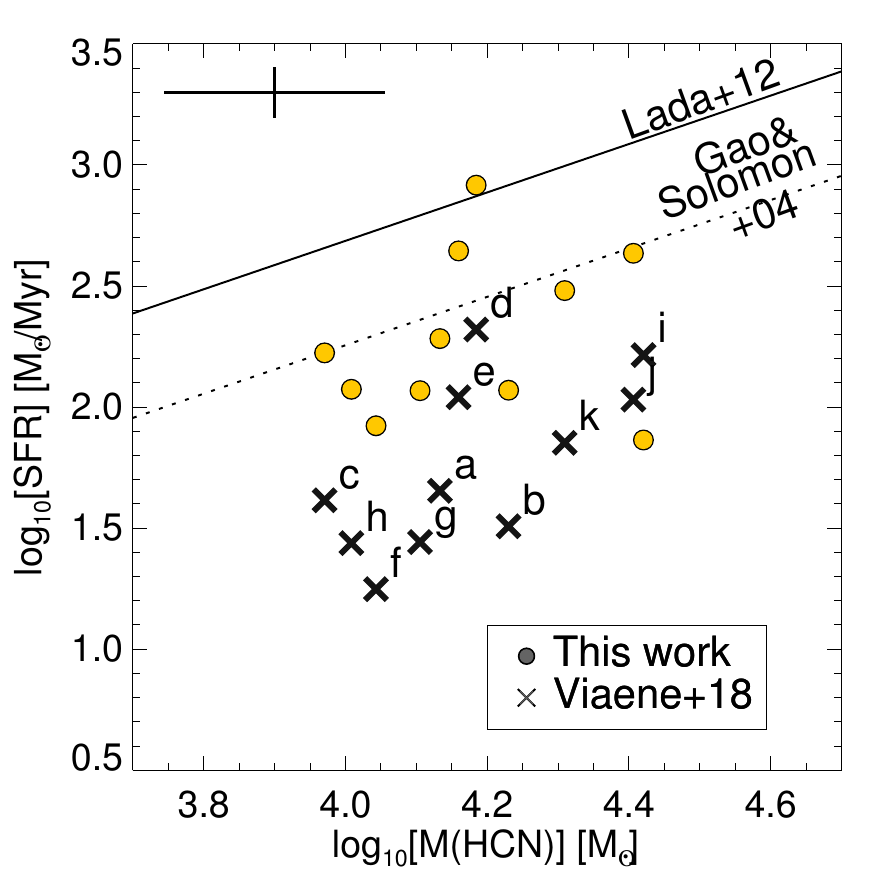}
\caption{  The SFRs as a function of the dense molecular gas (HCN) masses of the giant molecular clouds in M31. The masses of the molecular clouds are derived by \citet{Brouillet05}. The data from  \citet{Viaene18} (x symbols) have SFRs estimated from the SFR map of M31, created by \citet{Ford13}. Labels indicate names of the molecular clouds, as listed in \citet{Viaene18}. On the other hand, we estimated SFRs (circles) from the hybrid SFR(FUV+24$\mu$m) prescription from this paper (Appendix \ref{Sec:Appendix_B_Prescription}). The SFR values of the clouds predicted by  \citealt{Gao04} and  \citet{Lada12} are shown with dashed and solid lines, respectively. We show mean error bars of the data in upper left corner. }
\label{fig:Viaene}
\end{figure}

 There is an additional  evidence for the reliability of our reference SFR tracer. \citet{Viaene18} found that the giant molecular clouds in M31 exhibit $\approx$0.5 dex lower SFRs then what is predicted by Milky Way studies of the dense molecular gas (\citealt{Gao04,Lada12}). 
In their study, \citet{Viaene18} used the SFR map of M31  created by \citet{Ford13},  where the old stellar population contribution is subtracted.   
However, when we apply our hybrid SFR(FUV+24\,$\mu$m) prescription derived from  $\Sigma_{\rm SFR}$(H$\alpha,\rm corr$) (from Appendix \ref{Sec:Appendix_B_Prescription}), the SFRs of those molecular clouds match better with the values predicted by \citet{Gao04} than the SFRs used by \citet{Viaene18}. 
We show this on Fig. \ref{fig:Viaene}.

\section{Calibration of the SFR prescriptions}\label{Sec:SFR Calibrations}

In this section, we present the main results of our SFR calibrations. We compare $\Sigma_{SFR}$(H$ \alpha,\rm corr$) with $\Sigma_{SFR}$ derived from different  (monochromatic and hybrid) tracers. The comparisons are always shown with $\Sigma_{SFR}$(H$ \alpha,\rm corr$) on the x-axis, and $\Sigma_{SFR}$ from other tracers on the y-axis.  We also compare our SFR prescriptions with those of \citet{Calzetti07} and \citet{Leroy08}.  Hereafter, we will refer to \citet{Calzetti05} and \citet{Calzetti07}  as C05 and C07, respectively. Their prescriptions are similar to those given in \citet{Kennicutt03}, \citet{Leroy12} and \citet{CatalanTorrecilla15}. Moreover,  we evaluate the effects of varying the spatial resolution and subtracting the diffuse emission from non-star-forming regions on the SFR prescriptions. 

The monochromatic and hybrid SFR prescriptions at different resolutions and aperture sizes are listed in Tab. \ref{tab:AppB_Tab04} and \ref{tab:AppB_Tab05}  (Appendix \ref{Sec:Appendix_B_Prescription}).  We also include the monochromatic calibrations for 12\,$\mu$m, 70\,$\mu$m, 160\,$\mu$m and TIR tracers in Tab. \ref{tab:AppB_Tab05} (Appendix \ref{Sec:Appendix_B_Prescription}). In the same table, we add the SFR prescription for 12\,$\mu$m and 22\,$\mu$m calculated by fitting lines between the logarithmic values of L(IR) and SFR(H$\alpha$,corr), instead of surface densities.

\begin{figure*}[t!]
\centering
\includegraphics[width=0.9\linewidth]{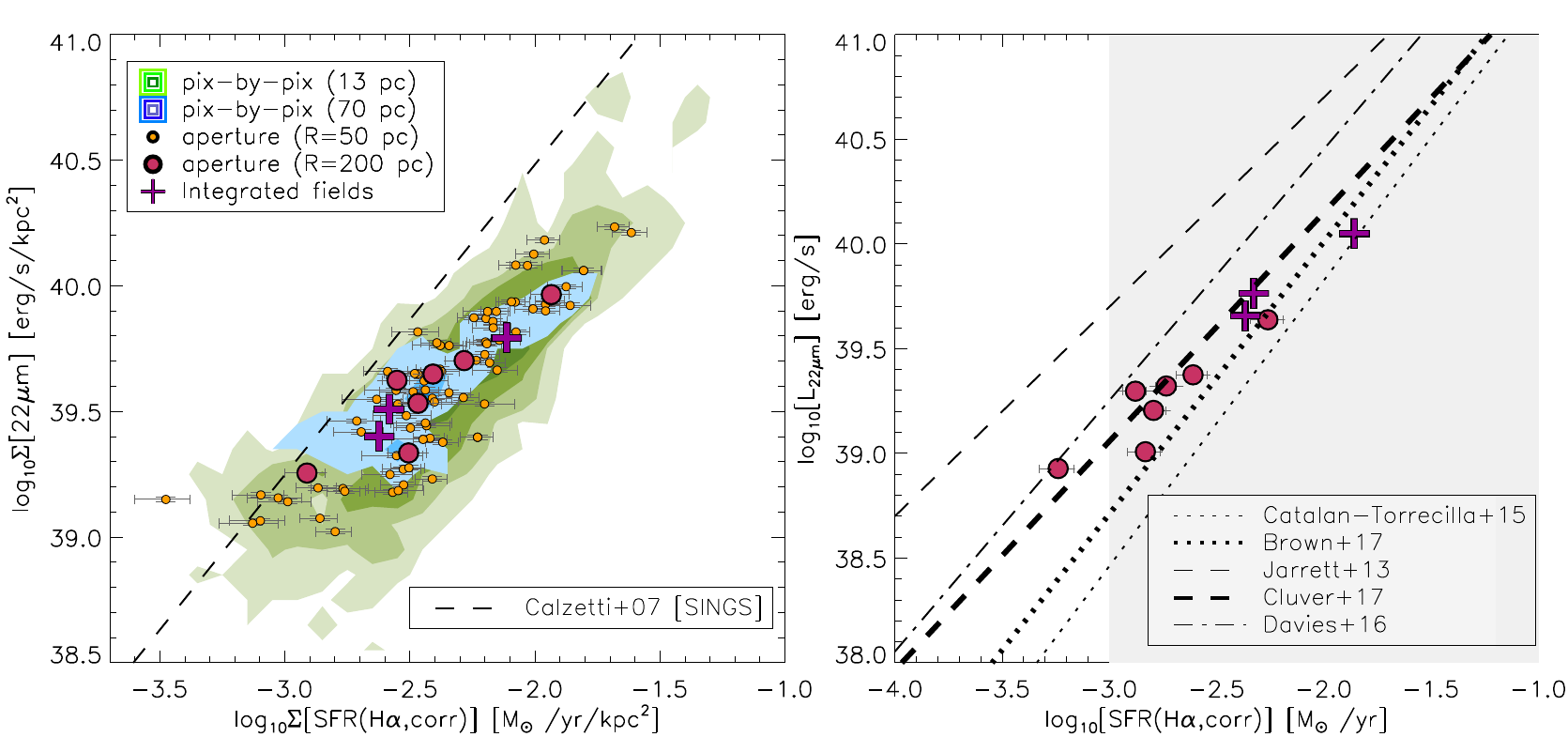}
\caption{ On the left, we show  $\rm \Sigma(\rm 22\mu \rm m) $  as a function of  $\rm \Sigma_{SFR}(\rm H\alpha,\rm corr) $, while on the right we show L$ (\rm 22\mu \rm m) $ as a function of SFR$\rm _{SFR}(\rm H\alpha,\rm corr)$. The M31 data presented here probe different spatial scales. Circles mark apertures, contours pixel-by-pixel data points, and crosses the integrated fields. Lines indicate relations and monochromatic SFR prescriptions given by  \citet[dashed line on the left panel]{Calzetti07}, \citet[dotted line]{CatalanTorrecilla15}, \citet[thick dotted line]{Brown17}, \citet[thin dashed line]{Jarrett13}, \citet[thick dashed line]{Cluver17}, and   \citet[dash-dotted line]{Davies16}. We indicate the range of SFRs covered by those papers with the shaded area. }
\label{fig:Mono}
\end{figure*}

\begin{figure*}[t]
\centering
\includegraphics[width=1.\linewidth]{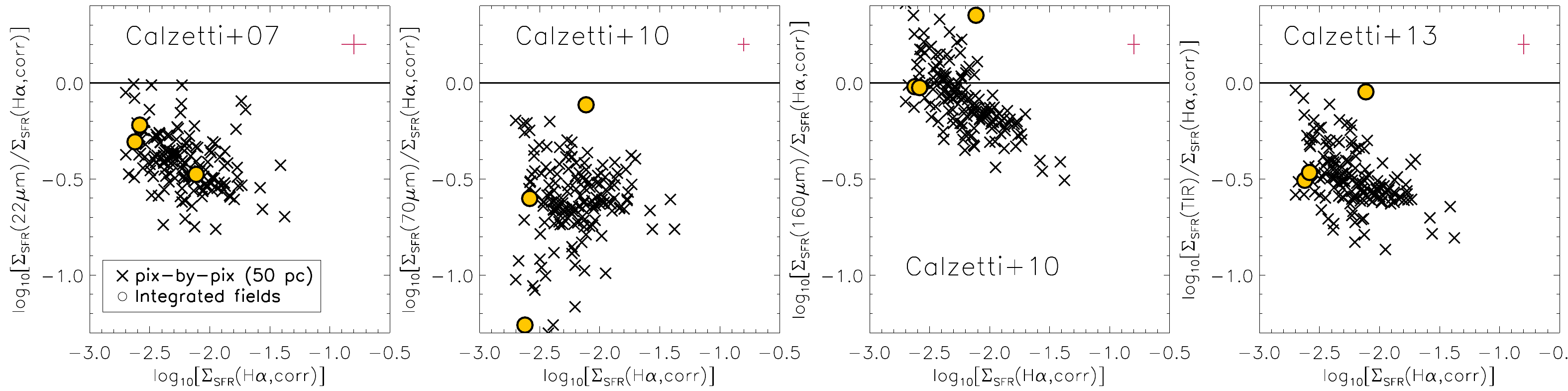}
\caption{ The ratio of different wavelength monochromatic $\rm \Sigma_{SFR}$(IR) calibrations given in the literature (\citealt{Calzetti07, Calzetti10, Calzetti13}) relative to  the  $\Sigma_{\rm SFR}$(H$\alpha,\rm corr$) as a function of $\Sigma_{\rm SFR}$(H$\alpha,\rm corr$). The tracers used here are the  $22\,\mu$m, $70\,\mu$m, $160\,\mu$m and TIR (estimated using Eq. 5 in \citealt{Dale09}) luminosities. The data points show pixel-by-pixel comparison (crosses; with 50\,pc pixel length) and integrated fields (yellow circles). The integrated field with the highest $\Sigma_{\rm SFR}$(H$\alpha,\rm corr$) is Field 3, a field that is mostly covered with HII regions. We show the mean error bars in the upper right corner (in red).   }
\label{fig:PACS_TIR}
\end{figure*}

\subsection{Monochromatic SFRs}  
\label{Subsec:Monochromatic}

The left panel of Fig. \ref{fig:Mono} shows the relation between  $\rm \Sigma_{SFR}$(H$ \alpha,\rm corr$) and monochromatic $\rm \Sigma(\rm 22\mu \rm m)$ at different pixel scales and apertures sizes. 
The dashed line indicates the monochromatic SFR prescription given by  \citet{Calzetti07} where they used apertures between 30\,pc and 1.2\,kpc in projected sizes. 
Here, the use of surface densities eliminates possible dependencies on spatial scales. 
Regardless of spatial scales, the M31 data show an 0.2-0.5 dex offset from the \citet{Calzetti07} prescription and a slope that is lower than 1.

On the right panel, we show  a comparison between the L(22$\rm \mu m$) and  SFR(H$ \alpha,\rm corr$).
We use here the luminosity and SFR values in order to facilitate the comparison of our data to monochromatic SFR  prescriptions from the literature, indicated by lines. 
The monochromatic SFR prescriptions from the literature are  provided by \citet{CatalanTorrecilla15}, \citet{Davies16}, \citet{Brown17}, \citet{Jarrett13}, and \citet{Cluver17}. 
These prescriptions were derived from extragalactic surveys with scales larger than 1 kpc and employed spectroscopic measurements and extinction corrected H$\alpha$. 
The exceptions are those from \citet{Jarrett13} and \citet{Cluver17}, where they used integrated galactic values of mid-IR photometry and  SFR measured from TIR. 
All these prescriptions are determined for SFR$ >10^{-3}\,\rm M_{\odot}\,yr^{-1} $ (represented by the gray shaded area in the figure), which is higher than the majority of our data (except for the integrated fields and the largest apertures). 
Therefore for consistency, we show only M31 data probing the largest scales.
Combining our largest apertures and the integrated fields, we observe that our data altogether are consistent with a single slope that is shallower than most of the monochromatic SFR prescriptions in the literature.  
These data fall between relations from \citet{CatalanTorrecilla15}, \citet{Cluver17}, and   \citet{Brown17}.

 Fig. \ref{fig:PACS_TIR} shows the residuals between $\Sigma_{\rm SFR}$(H$\alpha,\rm corr$) and $\Sigma_{\rm SFR}$(IR)   as a function of $\Sigma_{\rm SFR}$(H$\alpha,\rm corr$) for different IR tracers.
The IR tracers here are $22\,\mu$m, $70\,\mu$m, $160\,\mu$m and TIR luminosity. $\Sigma_{\rm SFR}$(IR) are derived from prescriptions given by \citet{Calzetti07}, \citet{Calzetti10} and \citet{Calzetti13}. 
The TIR values in this work are calculated using Eq. 5 in \citet{Dale09}, where we directly substitute the  $22\,\mu$m measurements for the $24\,\mu$m ones. 
We see that in M31 the $22\,\mu$m, $70\,\mu$m,  and TIR calibrations all under-predict the SFR relative to the H$\alpha$ one by $\approx0.5$\,dex with a 0.5\,dex scatter, while the $160\,\mu$m tracer under-predicts the SFR relative to the H$\alpha$ one by 0.1\,dex with a 0.5\,dex scatter.
However, we note that the  SFR($160\,\mu$m) prescription in the literature has a large uncertainty  as $160\,\mu$m is close to the peak of the IR SED and thus traces more the overall dust emission and is dominated by the cold dust emission. 
Note that for the TIR luminosity we assume a SF timescale of 100 Myr.

\begin{figure*}[t]
\centering
\includegraphics[width=0.8\linewidth]{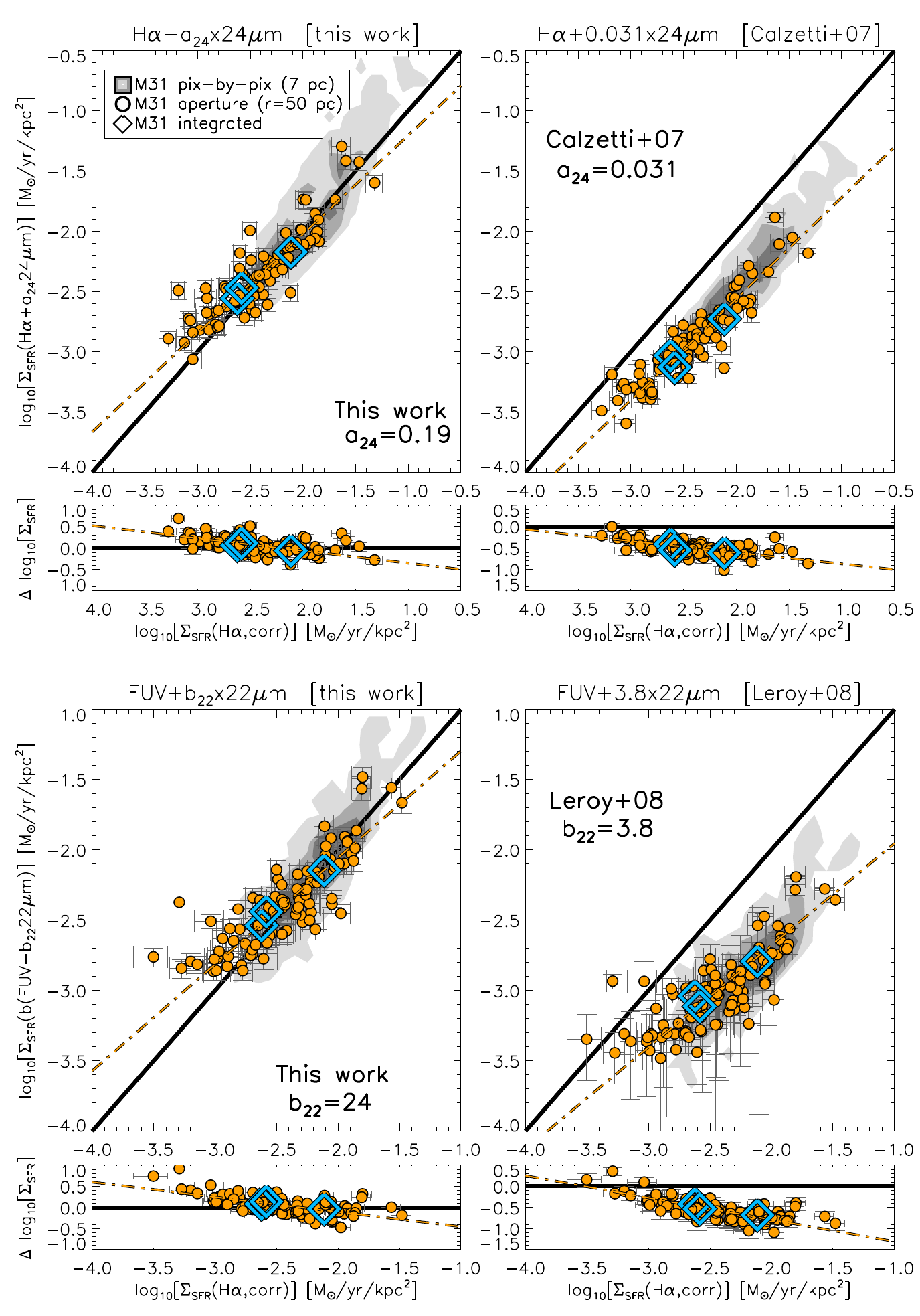}
\caption{ Comparison between hybrid $\Sigma_{\rm SFR}$ prescriptions and $\Sigma_{\rm SFR}$(H$\alpha,\rm corr$), using the hybrid SFR prescriptions from this work (left panels) or the prescriptions given by \citet{Calzetti07} and \citet{Leroy08} (right panels). We use different tracers for the hybrid SFRs: H$\alpha$+24$\,\mu$m (upper panels) and FUV+22$\,\mu$m (bottom panels). For the hybrid  SFR prescriptions, we use  H$\alpha,\rm corr$-SFR and FUV,corr-SFR (labeled as \textit{b} on axis) conversions factors from relations in Eq.  ~\ref{eq:Eq01} and   ~\ref{eq:Eq02}). Below each main panel, we plot residuals of the data from the above panel, where we calculate residuals as the hybrid SFR value subtracted from SFR(H$\alpha,\rm corr$). The contours show pixel-by-pixel data (7\,pc in size), the yellow circles R=50\,pc apertures and the blue diamonds the integrated fields (each with a projected size of $\rm \approx0.6\,kpc\times0.9\,kpc$).   In all panels, we plot the 1-1 relation (solid black line) that indicates an equivalence between the $\Sigma_{\rm SFR}$ values and  fits of the aperture data (in log-space as yellow dahsed-dotted lines). The prescriptions given by \citet{Calzetti07} and \citet{Leroy08} differ systematically from our work (seen as offsets between the data and the equivalence line on right panels).   }
\label{fig:Hybrid_vs_Hacorr}
\end{figure*}

\subsection{The hybrid SFR prescriptions at smallest scales} 
\label{Subsec:SFTprescriptions}

 We calibrate the SFR prescriptions for the hybrid tracers (H$ \alpha $+IR and FUV+IR)  by comparing them with the $\Sigma_{\rm SFR}$(H$\alpha,\rm corr$) at the smallest spatial scales, without subtraction of the diffuse emission. The smallest pixel-by-pixels scales of $\approx$7\,pc and the smallest aperture radius of $\approx$50\,pc are comparable to the HII region sizes in M31 that are between 15\,pc and 160\,pc (\citealt{Azimlu11}).

  We calculate  $ \Sigma_{SFR}(\rm H\alpha+IR) $ and $ \Sigma_{SFR}(\rm FUV+IR) $ as :
\begin{equation}
\Sigma_{\rm SFR}(\rm H\alpha+a_{IR}IR)=a\times[\rm \Sigma(\rm H\alpha_{observed})+a_{IR}\Sigma(\rm IR)],
\label{eq:Eq03} 
 \end{equation}
 \begin{equation}
\Sigma_{\rm SFR}(\rm FUV+b_{IR}IR)=b\times[\rm \Sigma(\rm FUV_{observed})+b_{IR}\Sigma(\rm IR)],
\label{eq:Eq04}
  \end{equation}
where mid-IR corresponds to 22\,$\mu$m and 24\,$\mu$m. The conversion factors \textit{a} and \textit{b} are 5.37$\times10^{-42}$ and  $4.42\times10^{-44}$, respectively, given from Eq. ~\ref{eq:Eq01} and ~\ref{eq:Eq02}.  The single-valued calibration factors a$_{\rm IR}$ and b$_{\rm IR}$ are used to account for obscured  emission of the tracers (H$ \alpha $ and FUV) and to recover extinction-corrected H$\alpha,\rm corr$ and FUV,corr. Those single-valued factors were measured by taking a median value of the calibration factors from individual data.  We calculate the factors  a$_{\rm IR}$ and b$_{\rm IR}$ for individual data as 
  \begin{equation}
  a_{\rm IR}=\dfrac{\Sigma(H\alpha,\rm corr)-\Sigma(H\alpha)}{\Sigma(\rm IR)},
  \label{eq:Eq for a}
  \end{equation}
 \begin{equation}
 b_{\rm IR}=\dfrac{\Sigma(\rm FUV,corr)-\Sigma(\rm FUV)}{\Sigma(\rm IR)}.
   \label{eq:Eq for b}
  \end{equation}
 
 The calibration factor a$_{\rm IR}$  is independent of the H$\alpha,\rm corr$-SFR conversion factor because a$_{\rm IR}$ is derived directly from observable tracers (H$\beta$, H$\alpha$ and IR).
 On the other hand, b$_{\rm IR}$ is sensitive to how we estimate SFRs, which depend on how we define the conversion factors \textit{a} and \textit{b}, and may differ  with different assumptions  taken in Sec. \ref{Subsec:HaToSFR}.

 In Fig. \ref{fig:Hybrid_vs_Hacorr}, we compare $\Sigma_{\rm SFR}$(H$\alpha,\rm corr$) with  the hybrid $\Sigma_{\rm SFR}$ calculated from our prescription (left panels) and from the prescriptions given by C07 and \citet{Leroy08} (right panels). 
 The SFR(H$\alpha$+24$\,\mu$m) and SFR(FUV+22$\,\mu$m) are presented in the upper and lower panels, respectively. 
The figure shows data from apertures with R=13.5'' ($\approx$50\,pc), pixel-by-pixel comparison (pixels with 7\,pc size) and the integrated fields.
 The residuals, presented below the main panels, show the difference between $\Sigma_{\rm SFR}$(H$\alpha,\rm corr$) and the hybrid $\Sigma_{\rm SFR}$ values.
 For all the panels, we also show the one-to-one relation and  power-law fits \footnote{We used  the \textit{IDL} tool mpfitexy for fitting (https://github.com/williamsmj/mpfitexy) including the estimated errors of the data. } for the aperture data.

All the panels in Fig. \ref{fig:Hybrid_vs_Hacorr} show a clear correlation between the hybrid $\Sigma$(SFR) and $\Sigma_{\rm SFR}$(H$\alpha,\rm corr$).
The scatter is usually between 0.3 and 0.5 dex. The right panels in Fig. \ref{fig:Hybrid_vs_Hacorr} shows clear systematic differences between the SFR values that are derived from H$\rm \alpha,corr $ and the SFR values derived from the prescriptions given by C07 and \citet{Leroy08}.
The discrepancy between those values is around 0.5 dex, and may be up to 1 dex. 
Our calibration leads to the calibration factors a$_{\rm 24}\approx0.2$ and b$_{\rm 22}\approx22$, which are about 5-8 times larger than those given by C07 and \citet{Leroy08} \footnote{a$_{\rm 24}\approx0.031$ in C07, a$_{\rm 24}\approx0.05$ in C05 and b$_{\rm 24}\approx3.8$ in \citet{Leroy08}.}.

\begin{figure*}[t]
\centering
\includegraphics[width=0.8\linewidth]{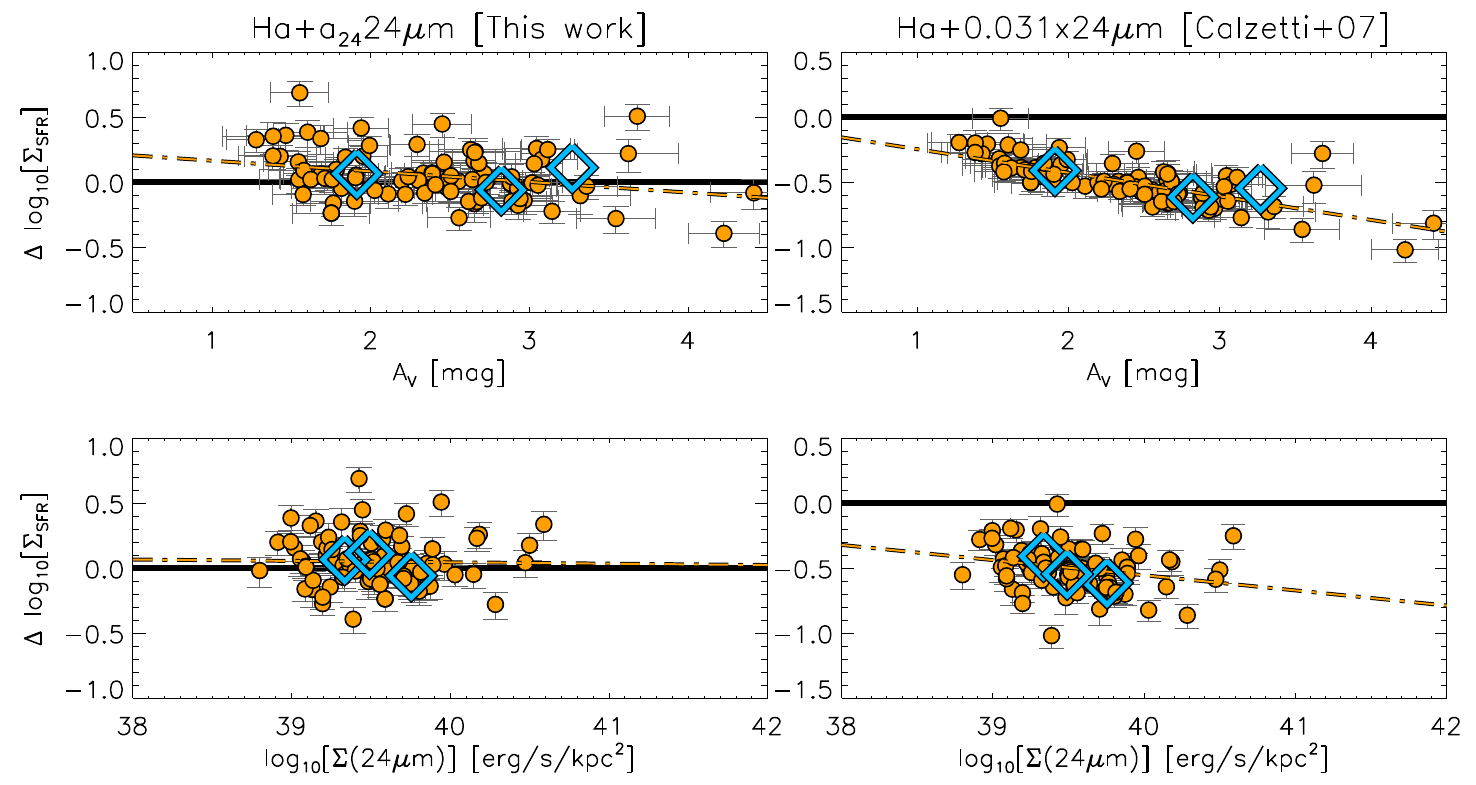}
\caption{We show here  residuals between $\Sigma_{\rm SFR}$(H$\alpha,\rm corr$)      and the hybrid H$\alpha$+24$\,\mu$m values,  as a function of physical quantities: A$_{\rm V}$ (upper panels) and  24$\,\mu$m  surface brightness (bottom  panels).  We show  apertures with radii of 50\,pc  (circles) and integrated fields (diamonds).  On the left, we show the behavior for the prescription from this work, while on the right panels we show the residuals for the C07 prescription (a$_{\rm 24}=0.031$).  We add linear fits (dashed lines) of the data (in logarithm) to the diagrams in order to better trace correlations between different quantities.     }
\label{fig:Hybrid_vs_Hacorr_res}
\end{figure*}

In Fig. \ref{fig:Hybrid_vs_Hacorr_res} we present the residuals as a function of two physical quantities: A$_{\rm V}$ (derived from the Balmer lines) and 24$\,\mu$m surface brightness. The residuals presented here are from the main panels of Fig. \ref{fig:Hybrid_vs_Hacorr} for the hybrid H$\alpha$+24$\,\mu$m prescriptions. Similarly, we show the results from this work on the left panels, and from using the C07 calibration on the right panels. Power-law fits are also included in the plots.

 In the top panels, the residuals do not change with A$_{\rm V}$ for our prescription (left panel). However, the residuals anti-correlate with A$_{\rm V}$ when using the C07 prescription (right panel). This can be easily explained by the low value of a$_{\rm 24\,\mu\rm m}$ from C07. The  $\Sigma$(24$\,\mu$m) values  are usually an order of magnitude higher than the observed $\Sigma$(H$\alpha$). When we multiply $\Sigma$(24$\,\mu$m) by a small a$_{\rm 24}$, the observed $\Sigma$(H$\alpha$) dominates over $\Sigma$(24$\,\mu$m). There is no clear trend in residuals with $\Sigma$(24$\,\mu$m) for our data, but a small trend when using the C07 prescription.

 We conclude that the SFR prescriptions at small spatial scales in M31 are different from those in the literature.  The $\Sigma_{\rm SFR}$(H$\alpha,\rm corr$) values are a factor of 3 ($\approx$0.5 dex) higher than the values obtained when using the prescriptions from the literature. Not only are the values different, but the scatter of the data is also large (0.3-0.5 dex).

\subsection{Effects of spatial scales and diffuse component subtraction}  
\label{Subsec:ScalesAndDiff}

 \begin{figure}[t!]
\centering
\includegraphics[width=0.8\linewidth]{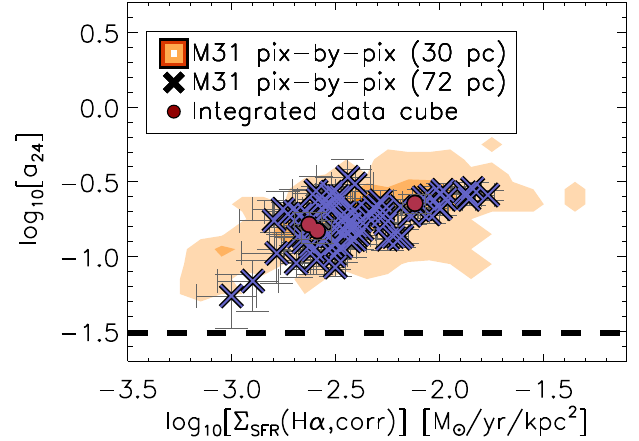}
\caption{ The effect of varying spatial scales on the calibration factor a$_{24\,\mu\rm m}$ as a function of $\Sigma$(H$\alpha,\rm corr$). The pixel-by-pixel data points shown in the  panel are for pixels at 25'' resolution (SPIRE 350$\,\mu$m, contours), 65'' resolution (blue X symbols) and the integrated fields (red circles). The data for  apertures with different radii have similar values as the presented data. We indicate the  a$_{24}=0.031$ value from C07 with the dashed line.     }
\label{fig:Scales}
\end{figure}

The prescriptions given in C07 and \citet{Leroy08} are derived from apertures and maps with lower spatial resolutions (C07 applied apertures with radii ranging from 0.03 kpc to 1.26 kpc, while \citealt{Leroy08} probe spatial scales at 800\,pc). They also included  procedures to subtract diffuse emission from mid-IR cirrus and DIG. Thus, to properly compare the prescriptions we need to test how the prescriptions vary when changing spatial scales and with a subtraction of the diffuse emission.

We show the effects of varying spatial scales on the calibration factor a$_{24}$ as a function of $\Sigma$(H$\alpha,\rm corr$), for the pixel-by-pixel based analyses in Fig. \ref{fig:Scales}. 
The C07 value of a$_{24}=0.031$  is presented as the dashed line. The difference between the C07 factor and  that from this work is around $\approx$0.7 dex at all spatial scales (maximum of 1 dex difference).
Our a$_{24}$ decreases from $\approx$0.22 at smallest scales to  $\approx$0.17 for the field-integrated measurements, and b$_{24}$ from $\approx$30 to  $\approx$20 (see Tab. \ref{tab:AppB_Tab04}).
We also indicate the integrated fields data  in Fig. \ref{fig:Mono}, \ref{fig:PACS_TIR}, \ref{fig:Hybrid_vs_Hacorr} and \ref{fig:Hybrid_vs_Hacorr_res} to show that, even at the same physical scales, the data in M31 consistently display  an offset in the monochromatic and  hybrid SFR prescriptions from the values given in the literature.

 \begin{figure}[t!]
\centering
\includegraphics[width=0.8\linewidth]{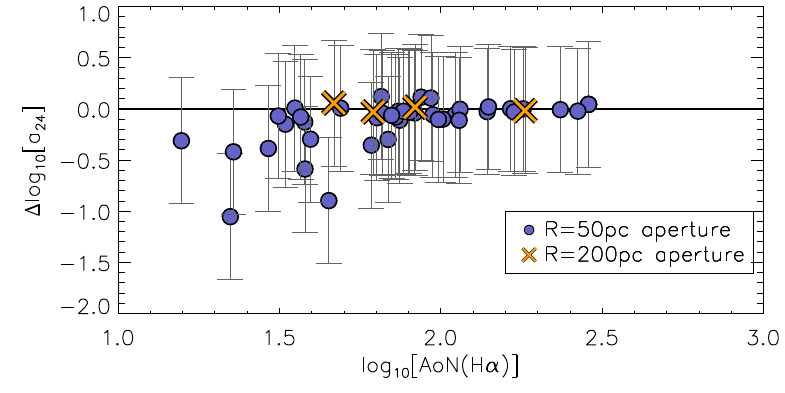}
\caption{ Difference in the a$_{24}$ values before and after subtracting the diffuse emission component (DIG and mid-IR cirrus) as a function of AoN(H$\alpha$). Apertures of  13.5'' radius (circles) and 55'' radius ($\times$ symbols) are presented. Unity is depicted with the solid line. Differences between the a$_{24}$ values from C07 and this work are usually 1 dex as seen in Fig.  \ref{fig:Scales}. We conclude that subtracting diffuse emission cannot explain the difference between the SFR prescriptions. The biggest impact of the diffuse subtraction is seen for  data of low AoN(H$\alpha$) or low surface brightness. }
\label{fig:Diffuse subtraction}
\end{figure}

Fig. \ref{fig:Diffuse subtraction} shows how subtracting diffuse emissions (DIG and mid-IR cirrus) affects the a$_{24}$ values. We use 13.5'' and 55''  radii apertures, and measure the mid-IR cirrus and DIG brightness by taking the mode of all pixels outside all of the  apertures in each M31 field respectively. The diffuse fraction of DIG and mid-IR cirrus in the apertures ranges from 5\% (for apertures with high surface brightness) to 30\%-60\% (for apertures with low  surface brightness).   After the subtraction, we see no change in a$_{24\,\mu\rm m}$ for the  high surface brightness (and high S/N) data. As expected, a stronger effect on a$_{24}$ is seen for the low surface brightness data.

\section{Effects of inclination and galactocentric radius on the SFR prescriptions}\label{Sec: aIR vs InclR25}

We compare the M31 data with observations from other galaxies to examine how attenuation,  galactic inclination and galactocentric distance affect the SFR prescriptions, and to place M31 in context with other galaxies. 
Variations in the individually estimated a$\rm _{24\mu m}$ factors are presented in Appendix \ref{Sec:a_vs_others} (see Fig. \ref{fig:a22_vs_others})
as a function of SFRs, IR emission and  dust temperatures in M31 and nearby galaxies.  

\begin{figure*}[t]
\centering
\includegraphics[width=1.\linewidth]{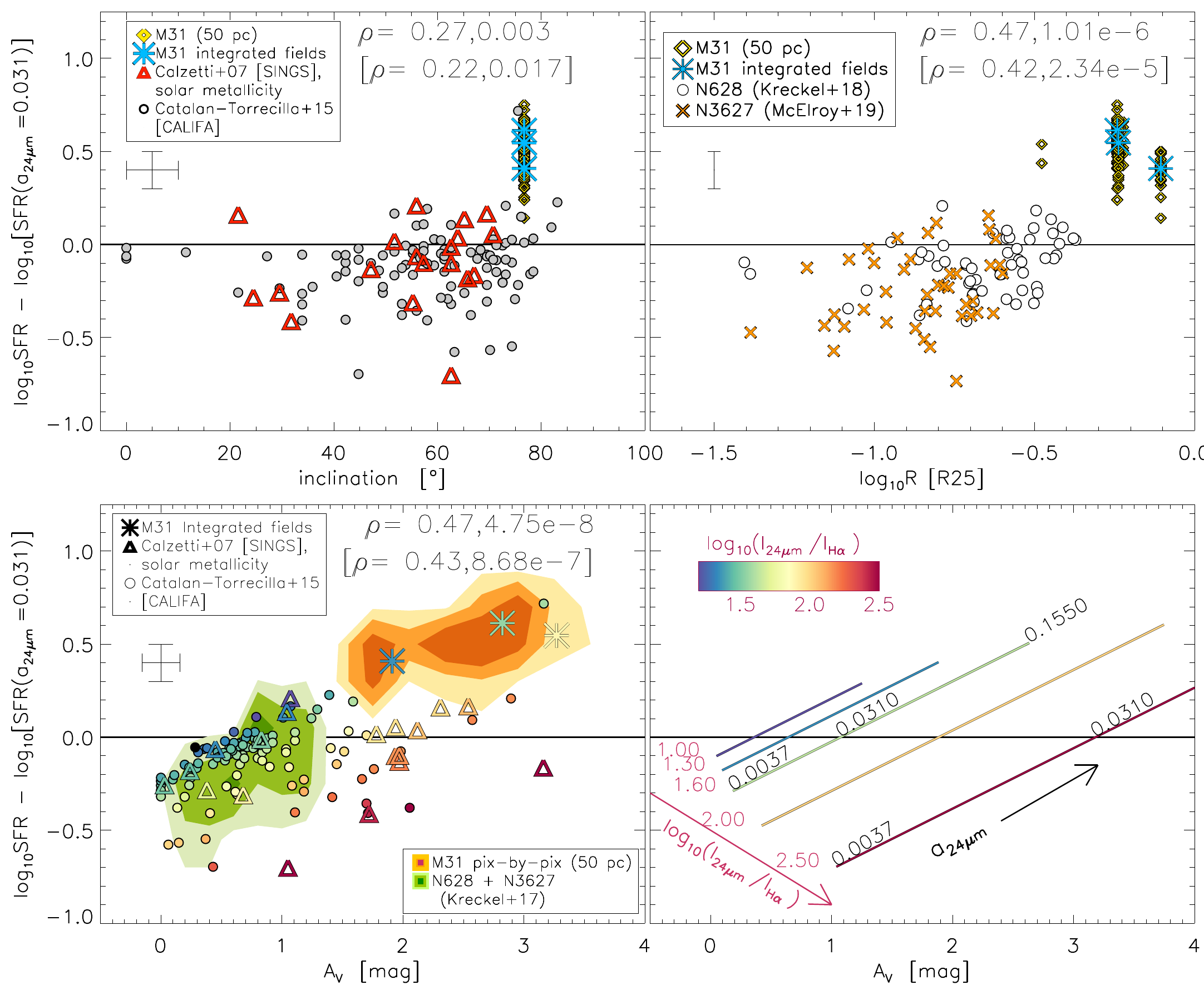}
\caption{Difference between SFR values estimated from the Balmer emission lines and SFR values from the hybrid SFR(H$\alpha$+a$\rm _{24\mu m}$24$\mu $m) prescription defined by C07, as a function of galactic inclination (upper left panel), galactocentric radius (upper right panel), and attenuation (lower panels). Each SFR value on these diagrams has been 
derived by calculating the corresponding a$\rm _{24\mu m}$ factor, thus the offset in SFR values reflects the difference between the estimated a$\rm _{24\mu m}$ factor and the factor from C07.
Data used are as follows: M31 (asterix for integrated fields; yellow contours or diamonds for the 50\,pc size pixels), SINGS sample of galaxies with metallicities similar to M31 (triangles; \citealt{Calzetti07}), CALIFA survey of galaxies (circles; \citealt{CatalanTorrecilla15}), and the MUSE data of NGC\,628 and NGC\,3627 (green contours in lower panel; circles and crosses in upper right panel; \citealt{Kreckel18,McElroy18inprep}). \textit{Upper panels --} Difference between SFR values as a function of galactic inclination (estimated using Eq. 1 in \citealt{Bergh88}) and the galactocentric distance (in units of R$_{25}$).   \textit{Lower panels --} Difference between SFR values as a function of attenuation for observed data points (left panel). The data are color-coded by their observed $\rm log_{10}(I_{24\mu m}/I_{H\alpha})$ ratio. In the lower right panel, we show the theoretically expected behaviour of the data assuming different $\rm log_{10}(I_{24\mu m}/I_{H\alpha})$ ratios and different a$\rm _{24\mu m}$ factors. The a$\rm _{24\mu m}$ values cover the range observed by \citet{Leroy12} (Fig. 9 in their paper). 
Uncertainties are shown on the left. The Spearman's  correlation coefficient ($ \rho $, left number) and  the significance of its deviation from zero (right number) are provided for all data points including M31 (upper numbers), and excluding M31 (numbers in brackets).   }
\label{fig:a_function_AinclinationR}
\end{figure*}

Fig. \ref{fig:a_function_AinclinationR} shows  differences between  SFR values estimated from the Balmer emission lines, and  SFR values estimated from the C07 prescription (assuming a$\rm _{24\mu m}=0.031$), as a function of galactic inclination, galactocentric radius and  observed  attenuation.
Each SFR value on these diagrams was derived using a$\rm _{24\mu m}$ factors that are individually estimated for each data point.
Thus the difference in the SFR values presented here  indicates a difference between the estimated a$\rm _{24\mu m}$ factors and the value of a$\rm _{24\mu m}=0.031$ from C07.

The data shown in the figure are from M31 (integrated fields and pixels with 50\,pc size), the SINGS galaxies (from C07), CALIFA\footnote{The Calar Alto Legacy Integral Field Area Survey; \citet{Sanchez12}} survey of galaxies from \citet{CatalanTorrecilla15}, and NGC\,628 and NGC\,3627 observed by the  MUSE\footnote{The Multi Unit Spectroscopic Explorer; \citet{Laurent06}} instrument (\citealt{Kreckel18,McElroy18inprep}).
Here, the data from the SINGS galaxies probe the central regions (out to galacto-centric distance between 0.5 and 2.5 kpc), for NGC\,628 and NGC\,3627  the data covers the central areas with binned pixel of 0.3 kpc and 0.8 kpc sizes respectively, and the data from  the CALIFA survey which are from apertures (with 36'' radius) covering entire galaxies or most of their (optical) disks. 
The M31 pixels data are binned to 50\,pc (driven by the WISE resolution) in order to plot spatially independent pixels. 
Additionally, we show the Spearman's  correlation coefficient  and  the significance of its deviation from zero. 
This coefficient is given for all data including and excluding the integrated M31 fields, in order to highlight correlations among galaxies even when excluding M31 data.

In the bottom right panel of Fig. \ref{fig:a_function_AinclinationR}, we present the expected theoretical trends of data for various SFR prescriptions (y-axis)  as a function of attenuation (x-axis),  $\rm log_{10}(I_{24\mu m}/I_{H\alpha})$ ratios (color), and different  a$\rm _{24\mu m}$ factors.  
We calculate the behaviour of these trends  taking the following steps.
Using the Eq. 1 and 8 in \citet{Calzetti94}, and the relation $\rm E_{B-V}\equiv A_V/R_V$, we derive the following relation between the observed H$\alpha$ and H$\rm \alpha,corr$:
\begin{equation}
 \rm H\alpha^{corr}=H\alpha^{obs}e^{0.921\cdot k_{\lambda}\cdot (A_V/R_V)} =H\alpha^{obs}10^{0.4\cdot k_{\lambda}\cdot (A_V/R_V)} ,
   \label{eq:Theoretical model p1}
  \end{equation}
  where IR corresponds to the 24\,$\mu$m emission, and  the $\rm k_{\lambda}=2.5186$ is the extinction value of the H$\alpha$ line, assuming the \citet{Cardelli89} extinction curve and R$\rm _V$=3.1. 
We then combine this relation with the Eq. \ref{eq:Eq for a} from this work, to derive the trends  on the x-axis of the diagram:
\begin{equation}
 \rm A_V=\frac{R_V}{0.4\cdot k_{\lambda}} \cdot log_{10} \big( 1+ a_{IR}\frac{I_{IR}}{I_{H\alpha} }  \big).
   \label{eq:Theoretical model p2}
 \end{equation}
For the trends  on the y-axis, we derive the following relation using Eq. \ref{eq:Eq01} from this work: 
\begin{equation}
 \rm log_{10}(SFR)=log_{10}(a) + log_{10}(H\alpha^{obs}) + log_{10}\big( 1+ a_{IR}\frac{I_{IR}}{I_{H\alpha} }  \big),
    \label{eq:Theoretical model p3}
 \end{equation}
 \begin{equation}
  \rm \Delta log_{10}[SFR]=log_{10}\big( 1+ a_{IR}\frac{I_{IR}}{I_{H\alpha} }  \big)-log_{10}\big( 1+ 0.031\frac{I_{IR}}{I_{H\alpha} }  \big).
   \label{eq:Theoretical model p4}
 \end{equation}
The range covered by the a$\rm _{IR}$ factor in these equations is the same as that observed by \citet{Leroy12} in nearby galaxies (their Fig. 9).

 Regarding the inlcination and galactocentric distance, the SFR prescription in Fig. \ref{fig:a_function_AinclinationR} shows  some correlation with galactocentric distance and a weak correlation with galactic inclination. 
Those correlations are still present, although slightly reduced, even when we exclude the M31 integrated fields. 
The scatter is large for inclination, which explains a lower Spearman's  correlation coefficient than in the case of the galactocentric distances. 
However, most of the nearby galaxies have inclinations lower than M31. 
The few galaxies with inclinations very similar to M31 show in general slightly higher a$\rm _{IR}$ factors compared to the other galaxies.  
For galactocentric distance, we find a slightly stronger trend, and a lower scatter for the estimated a$\rm _{IR}$ factors.
Given the lack of data around galactocentric distances of $\rm log(R/R_{25})\approx-0.65$, this trend is tentative.

The high inclination of M31 and the fact that our data probe large galactocentric radii may be a cause of the high attenuation values.  
This is supported by the fact that the SFR prescription and its offset from the C07 prescription correlates with attenuation. 
Furthermore, the M31 data points (integral or pixel-by-pixel) follow the trend seen in other nearby galaxies, and their observed   $\rm log_{10}(I_{24\mu m}/I_{H\alpha})$  ratios are consistent with
the expected trends in the model diagram. 
Variations in the $\rm log_{10}(I_{24\mu m}/I_{H\alpha})$ ratio increase the scatter in the SFR prescriptions for a fixed attenuation value. 
For example, data with the same a$\rm _{IR}$ factor exhibits higher attenuation values when increasing the $\rm log_{10}(I_{24\mu m}/I_{H\alpha})$ ratio.

In Fig. \ref{fig:a22_vs_others} (Appendix \ref{Sec:a_vs_others}), we note a trend of decreasing  a$\rm _{22\mu m}$ factor with increasing IR emission, observed H$\alpha$,  and the dust temperature. 
The M31 data tend to show higher a$\rm _{22\mu m}$,  lower IR emission, lower $70\,\mu\rm m/160\,\mu\rm m$ ratio (indicating colder dust), and  $160\,\mu\rm m/TIR$ ratio (indicating a higher fraction of the cold dust emission within the TIR)  compared to nearby galaxies from the CALIFA and SINGS surveys. 
We attribute the lower IR emission and lower observed H$\alpha$ in the M31 fields due to the fact that the M31 data probe large galactocentric distance. 
Thus the cold dust component observed at 160$\mu$m in the M31 fields has an additional contribution along the line-of-sight that arises from dust not heated by star formation. 
Given the high inclination and large galactocentric distances probed (Fig. \ref{fig:a22_vs_others}), a flaring dust disk (analogous to a flaring cold gas disks) would provide the best explanation for this behavior.

\section{Discussion}\label{Sec:Discussion}

We find  that both monochromatic and hybrid SFR prescriptions for our M31 fields, calibrated by using the   extinction corrected   $\Sigma_{\rm SFR}$(H$\alpha,\rm corr$), deviate from the standard prescriptions in the literature. 
The M31 fields yield values 5--8 times higher for the a$_{IR}$ and b$_{IR}$ factors correcting the observed H$\alpha$ or FUV emission in hybrid prescriptions, compared to the literature \citep{Calzetti05,Calzetti07,Leroy08,CatalanTorrecilla15}. 
In this section, we discuss what may cause this offset in the SFR prescriptions, and why the hybrid SFR prescriptions may not be universal, as assumed in the literature.

\subsection{Galactocentric radius and inclination}

 In the literature, hybrid SFR prescriptions are assumed to be universal, and to trace the `local' attenuation. 
This `local' attenuation of the light arising from star-forming regions is due to dust that surrounds the star-forming regions. 
Heated by the ionizing photons escaping from the star-forming regions, this dust is heated and emits in the mid-IR.
Hence mid-IR emission can be associated with attenuated SFR tracers (FUV and H$\alpha$) and be used for calibrating SFR prescriptions. 
This model also implies that there is no additional, more (vertically) extended dust component present in galaxies. 
If this model is correct, then we expect to see
no variations in the SFR prescriptions as a function of attenuation, inclination or galactocentric radius. 

However, comparison of data from galaxies and data within galaxies
reveals tentative to obvious trends for 
the SFR prescriptions as a function of inclination, galactocentric radii and attenuation (Fig. \ref{fig:a_function_AinclinationR}).
Further previous observations of nearby galaxies indicate that hybrid SFR prescriptions vary with specific quantities or with galactocentric distance. 
Firstly, \citet{Gonzales06} report a $\approx0.2$ dex offset in SFR(IR) values when comparing their results on the spiral arms in M81 with the M51 data from C05. They also observed lower $24\,\mu\rm m$/TIR ratios compared to the prediction from C05 (Fig. 4 in \citealt{Gonzales06}), which indicates that the dust in the M81 data is cooler, similar to what we see in M31.
Secondly, \citet{CatalanTorrecilla15} saw a weak correlation between observed attenuation and a$_{\rm IR}$ in their sample of CALIFA galaxies, as we do in Fig. \ref{fig:a_function_AinclinationR}.  
Thirdly, \citet{Boquien16} found that the b$_{\rm IR}$ factor increases from the centers toward the disk outskirts in their sample of face-on galaxies (Fig. 4 in their paper). 
They concluded that the b$_{\rm IR}$ factor and the SFR prescription change due to $\Sigma$(M$_{\rm stellar}$) and $\Sigma$(sSFR), which decrease toward the outskirts of galaxies. 
However, we argue that $\Sigma$(M$_{\rm stellar}$) and $\Sigma$(sSFR) alone can not explain the variations in the SFR prescriptions, as we estimate b$_{\rm IR}\approx9\pm2$ for our M31 fields when we apply their b$_{\rm IR}$-$\Sigma$(sSFR) conversion.
That value is still 2 times lower than the number determined from our calibration.

Analyzing the same M31 fields, \citet{Tomicic17} concluded that the vertical distribution of the dust and ionized gas must change differently as a function of galactocentric distance in order to explain why the measured attenuation in the outskirts of M31 is higher than the measured attenuation in the central regions of nearby galaxies.
If we assume that the dust is well mixed with the HI and H$_2$ gas (\citealt{Holwerda12,Hughes14}), the vertical scale-height of the dust should increase with a galactocentric radius as both HI and H$_2$ gas layers are thickening towards the outer disk (as seen for the highly inclined M31 and other galaxies; \citealt{Braun91,Olling96,Yim14}). 
On the other hand, the vertical scale height and luminosity of the ionized gas (DIG and HII regions) correlate with the number and intensity of star-forming regions, and decrease with the galactocentric radius (\citealt{Dettmar90,Rand96,Oey07,Bigiel10}). 

Therefore, we caution against the use of SFR prescriptions at large  galactocentric distances and in galaxies with high inclinations.
If the dust in the HI-dominated outskirts of galaxies is more extended along the line-of-sight and has no longer a close spatial association with star-forming regions, then it cannot easily be heated by ionized photons originating from these regions.
This geometry leads to higher attenuation that is not followed by higher mid-IR emission, and thus the a$\rm_{IR}$ and b$\rm_{IR}$ factors increase, changing the SFR prescriptions.
High galactic inclination also adds dust along the line-of-sight, which is not associated with star-forming regions, thus increasing the attenuation.
This may be the reason why on Fig. \ref{fig:a_function_AinclinationR} we see a variation in the SFR prescription as a function of galactocentric radius and inclination, which leads to the correlation with attenuation.  

M31 is a highly inclined galaxy, and the fields used in this paper probe galactocentric radii that are larger than the galaxies (KINGFISH, SINGS and CALIFA surveys, and NGC 3627, NGC 628) used for calibration of the SFR prescriptions in the literature. 
This leads to higher attenuation for the same dust mass surface density compared to nearby galaxies (\citealt{Tomicic17}), and larger a$\rm_{IR}$ and b$\rm_{IR}$ factors.
Due to the large galactocentric distance, the M31 field also have a low surface brightness in the mid-IR and observed Halpha emission (Fig. \ref{fig:a22_vs_others}). 
Moreover, this additional dust along the line-of-sight in M31 is not heated by star-forming regions, and thus has 160\,$\mu$m emission which dominates the total IR emission  and a  slightly  lower dust temperatures  (see Fig. \ref{fig:a22_vs_others}).

This is backed by the analysis of \citet{Groves12}, who found relatively cold dust in M31, and concluded that the dust is predominantly heated by the older stellar population. 
Similarly, by deriving a model of the dust heating in M31, \citet{Viaene17} also concluded that most of the radiation heating of the dust in the disk of M31  comes from an older stellar population (particularly in the bulge).
 \citet{Xu96} detected low $60\,\mu$m/$100\,\mu$m ratios in the M31 spiral arms, and concluded that the dust in M31 is cooler, more diffuse or lacking very small grains.

 \subsection{Spatial scales, age of the clusters and sampling initial mass function}

 As smaller  and smaller spatial scales are probed, standard SFR prescriptions can break down due to three main effects: 1)  SFR tracers that use reprocessed emission, such as H$\alpha$ and IR, may arise from star formation activity outside the region probed (e.g. from neighboring pixels; \citealt{Boquien16}), 2) the simple assumption of continuous star formation over 10-100\,Myr may break down, with stochastic sampling of stellar ages (i.e. individual single-aged clusters), 3) the initial mass function (IMF) is  no longer sampled fully, with stochastic sampling of high mass stars, and thus changing the assumptions made for Eq. \ref{eq:Eq01}. 
 The impact of moving to small spatial scales on the SFR prescription has been discussed by \citet{Faesi14}. They observed star forming regions in NGC\,300 at 250\,pc  scales, and used STARBURST99 \citep{Leitherer99} modeling  (assuming an instantaneous burst of star formation) to infer the SFR.
Their inferred SFRs are 2-3 times ($\approx$0.3 dex) higher compared to  the SFRs  assuming continuous star formation as used in  C07 and \citet{Leroy08}.
The authors argue that on small spatial scales where individual HII regions are observed, the burst model is more appropriate compared to measurements done on larger spatial scales where averaging within the aperture would correspond to a more uniform star formation history.
Similarly, \citealt{daSilva14} and \citet{Krumholz15} applying the SLUG\footnote{SLUG is a code that Stochastically Lights Up Galaxies, to simulate galaxies undergoing stochastic star formation.} software report stochastic fluctuations in star formation that can produce non-trivial errors for SFRs with biases of $>$0.5 dex at the lowest SFR values.

 In our case, integration of the entire M31 fields (covering 1-2 kpc in scale) probes all HII regions in this area ensuring that we capture all relevant emission and that the measurements are robust against stochastic variations in the age. 
Interestingly, integrated fields still show similar SFR prescriptions as those obtained at small scales.
Further our inferred calibration of the hybrid SFR(H$\alpha$+a$_{IR}$IR) prescriptions is not affected by age variations (sometimes plaguing the conversion of FUV to H$\alpha$ emission) as it only uses mid-IR, H$\alpha$ and H$\beta$ emission.
  Finally, \citealt{daSilva14} and \citet{Krumholz15} emphasize that the bias in the estimated SFR is most evident in regions with very low SFRs ($<5\,\rm M_\odot \,yr^{-1}$). 
Our apertures, integrated fields and larger pixels used for M31 fields probe a SFR range of $\rm log(SFR/M_\odot \,yr^{-1}\approx 4-2)$ where the SLUG simulations reveal no strong bias for the estimated SFR, but show a large 0.5 dex scatter (e.g. Fig. 6 in \citealt{daSilva14}).
Furthermore the spatial scales probed in the M31 fields are similar in size to the scales used by C07, \citet{Leroy12} and \citet{CatalanTorrecilla15}, and thus their prescriptions would suffer the same issues and biases. 
 Nonetheless, our  SFR prescriptions differ from the literature values, even when using similar spatial resolution. 
Note that the scatter of SFR values within our fields (seen in Fig. \ref{fig:Hybrid_vs_Hacorr}) and the scatter of residuals between our SFRs and the SFRs by \citet{Lewis15} (as seen in Fig. \ref{fig:Lewis}) decrease with increasing spatial scales, as predicted by  \citealt{daSilva14}.

\subsection{Effects of the diffuse emission components}

Our SFR prescriptions and a$_{IR}$ derived for M\,31 remain similar even
when we subtract diffuse emission, with only the lowest surface brightness regions being a exception (Sec. \ref{Subsec:ScalesAndDiff}). 
As we try to follow procedure from C07 for subtracting diffuse emission, we can only measure diffuse emission inside the M31 fields that are within the spiral arms unlike C07 that integrate areas outside the apertures and spiral arms to calculate the amount of diffuse emission.   
If our method of probing diffuse emission within the spiral arms results in too bright diffuse emission, the a$_{IR}$ obtained would be even higher compared to those not taking into account the diffuse emission. 
 
On the other hand, our diffuse fractions of 5\%-60\% are consistent with the typical numbers seen in nearby galaxies (\citealt{Leroy12}).
 Additionally, \citet{Leroy12} measure and subtract the cirrus using the dust emissivity throughout their galaxies, and still find a SFR prescription that differs from ours.

\subsection{Implications}

The first direct implication of our findings is that the traditional hybrid SFR prescriptions cannot be applied to M31 (or to be more precise the star-forming regions in its outer disk), and that previously estimated SFR values for star-forming regions in M31 are highly (around 0.5 dex, and up to 1 dex) unreliable using standard prescriptions from the literature. 
For example, in Sec. \ref{Subsec:Viaene} we show that the SFR values of molecular clouds in M31 are higher than those derived from the prescriptions commonly used. Our higher SFR values move these clouds back onto the relation between molecular gas mass and SFR (\citealt{Gao04}). 

A second corollary of our analysis is that the standard SFR prescriptions are not universally applicable even in local galaxies. 
In particular, they should be used with caution at large galactocentric distances (with higher a$_{\rm IR}$ and b$_{\rm IR}$  in the outskirts of galaxy disks) and in highly inclined galaxies.
Our results  imply that higher   a$_{\rm IR}$ and b$_{\rm IR}$ factors should be employed in the outskirts of galaxies.

\section{Summary}\label{Sec:Summary}

 In this paper, we calibrate  SFR prescriptions using  different tracers and considering different  spatial scales (between $\approx$10\,pc and $\approx$0.9\,kpc). 
We utilize high angular resolution observations available for five 0.6\,kpc$\times$0.9\,kpc fields in the spiral arms of M31: H$\alpha$ (from IFU data), 22$\,\mu$m (from WISE), 24$\,\mu$m (from SPITZER/MIPS) and FUV (GALEX). 
We also calibrated 12$\,\mu$m, 70$\,\mu$m, 160$\,\mu$m and TIR SFR prescriptions. 
Our reference SFR tracer is extinction corrected H$\alpha$ (H$\alpha,\rm corr$, using the Balmer decrement) which we compare with hybrid SFR tracers (H$\alpha$+IR or FUV+IR) and other (mono-chromatic) SFR prescriptions.

 Our main results can be summarized as follows: 
 
\begin{itemize}
\item $\Sigma_{\rm SFR}$(H$\alpha$,corr) agrees relatively well with $\Sigma$(SFR) derived from 
the modeled star formation history of M31 by  \citet{Lewis15}. Similarly, applying our SFR prescription to molecular clouds in M31 moves them onto the relation found by \citet{Gao04}, unlike SFRs estimates using standard prescriptions from the literature.

 \item The calibration factors (a$_{IR}$ and b$_{IR}$) for hybrid SFR prescriptions  are systematically a factor of 5-8 times larger in M\,31 than the ones stated in the literature \citealt{Calzetti07,Leroy08,CatalanTorrecilla15}). Similarly, our SFR(H$\alpha,\rm corr$) values in M\,31 are higher  than SFRs given by other prescriptions (0.5 dex). 
 
\item The SFR prescriptions (in M\,31) do \textit{not} change with spatial scales. Moreover, the subtraction of a diffuse component (neither DIG nor mid-IR cirrus) has no effect on the obtained prescription from our fields, except for slight variations in the lowest surface brightness (SFR$\rm<3 M_\odot\,yr^{-1}\,kpc^{-2}$) regions. 

\item  Compared to nearby galaxies used for calibrating the SFR prescriptions in the literature, the  M31 fields probe significantly larger galactocentric distances (by 3 times), high galactic inclination, and an order of magnitude lower mid-IR and H$\alpha$ surface brightness. The M31 fields also exhibit an on average lower $70\,\mu\rm m/160\,\mu\rm m$ ratio and a higher $160\,\mu\rm m$/TIR ratio (than the nearby galaxies), which indicate the presence of colder dust. We see evidence that the commonly used SFR prescriptions correlate with galactocentric distance, galactic inclination and attenuation.

\end{itemize}
We interpret these findings as follows:
\begin{itemize}
\item We propose that the SFR prescriptions are sensitive to variations in the relative (3-dimensional) dust/gas distributions across the galactic disks which change with inclination and galactocentric distance. Lines-of-sight towards outer galaxy disks where the dust/gas distribution has a larger scale height than towards 
galactic centers, or through more inclined galaxies, probe additional dust that is related
to HII regions and the diffuse gas. 
This dust layer is not directly associated with star-forming regions which results in a lower mid-IR surface brightness (dominated by mid-IR cirrus emission), colder dust, and higher A$_{\rm V}$ of the H$\alpha$ photons, and could explain the change we observe in the SFR prescriptions.  
 This view is consistent with recent results by \citet{Tomicic17} who showed that M31 Fields probed in this work have a different relative dust/gas distribution along the line of sight compared to other nearby galaxies. 

\end{itemize}

The authors wish to kindly thank Alexia Lewis, who shared the maps of modeled FUV emission and star formation history in M31, and to Sarah Leslie for additional comments. Tomi\v{c}i\'c  N. and Kreckel K. acknowledge grants SCHI 536/8-2 and KR 4598/1-2 from the DFG Priority Program 1573. This work is based on observations collected at the Centro Astron\`omico Hispano Alem\`an (CAHA), operated jointly by the Max-Planck Institut f\"ur Astronomie and the Instituto de Astrofisica de Andalucia (CSIC), and is also based on observations made with Herschel. Herschel is an ESA space observatory with science instruments provided by European-led Principal Investigator consortia and with important participation from NASA. PACS has been developed by a consortium of institutes led by MPE (Germany) and including UVIE (Austria); KU Leuven, CSL, IMEC (Belgium); CEA, LAM (France); MPIA (Germany); INAFIFSI/OAA/OAP/OAT, LENS, SISSA (Italy); and IAC (Spain). This development has been supported by the funding agencies BMVIT (Austria), ESA-PRODEX (Belgium), CEA/CNES (France), DLR (Germany), ASI/INAF (Italy), and CICYT/MCYT (Spain). SPIRE has been developed by a consortium of institutes led by Cardiff University (UK) and including Univ. Lethbridge (Canada); NAOC (China); CEA, LAM (France);
IFSI, Univ. Padua (Italy); IAC (Spain); Stockholm Observatory (Sweden); Imperial College London, RAL, UCL-MSSL, UKATC, Univ. Sussex (UK); and Caltech, JPL, NHSC, Univ. Colorado (USA). This development has been supported by national funding agencies: CSA (Canada); NAOC (China); CEA, CNES, CNRS (France); ASI (Italy); MCINN (Spain); SNSB (Sweden); STFC (UK); and NASA (USA). Based on observations made with the NASA Galaxy Evolution Explorer. GALEX is operated for NASA by the California Institute of Technology under NASA contract NAS5-98034.  This publication makes use of data products from the Wide-field Infrared Survey Explorer \citep{2010AJ....140.1868W}, which is a joint project of the University of California, Los Angeles, and the Jet Propulsion Laboratory/California Institute of Technology, funded by the National Aeronautics and Space Administration.

\bibliographystyle{yahapj}


\clearpage
\newpage

\appendix
\renewcommand\thefigure{\thesection.\arabic{figure}}

\section{Maps of the SFR tracers}
\label{Sec:Appendix_A_Maps}

\begin{figure}[t!]
\centering
\includegraphics[width=1.0\linewidth]{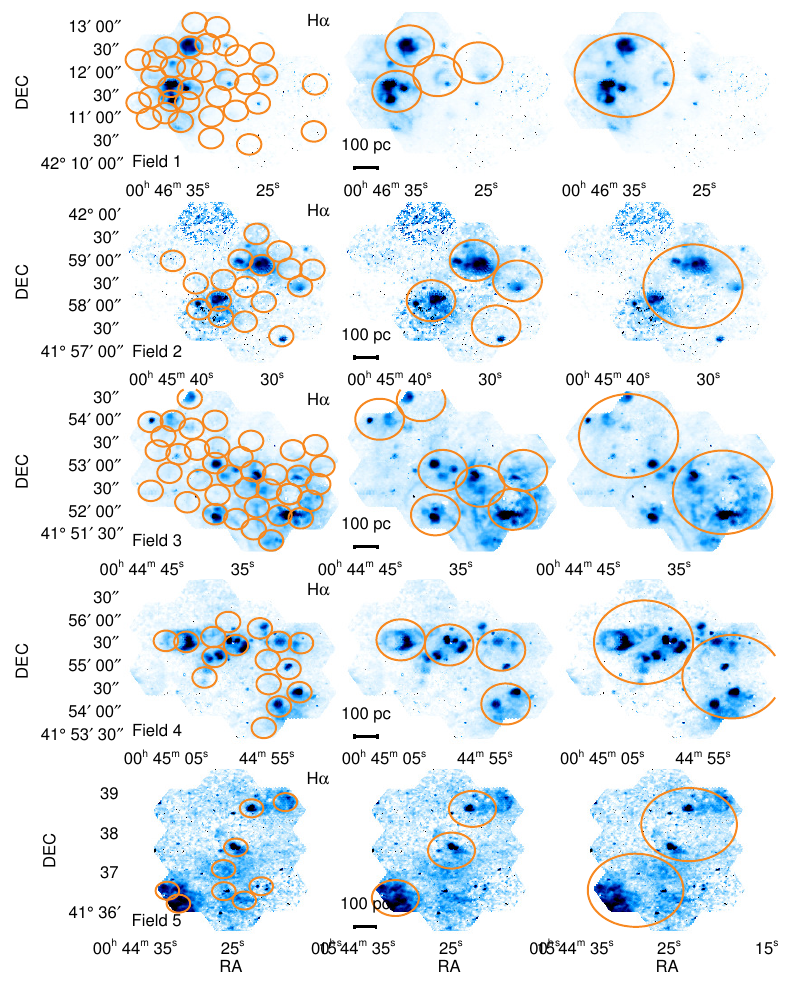}
\caption{ Maps of the  H$\alpha$ tracer for Fields 1 to 5 (from top to bottom), with overplotted apertures  with radii of 13''(50\,pc, left), 27''(100\,pc, center), and 55''(200\,pc, right).  
\label{fig:Apertures_fields}}
\end{figure}

 In Fig. \ref{fig:Apertures_fields}, we show the position of the apertures used in this work, overplotted on the H$\alpha$ maps. Apertures  have radii of 13''(50\,pc), 27''(100\,pc), and 55''(200\,pc).  Fluxes in the apertures were calculated using the \textit{IDL} software tool \textit{aper} to extract pixel values without additional interpolation of pixel fluxes.

\begin{figure}[t!]
\centering
\includegraphics[width=1.0\linewidth]{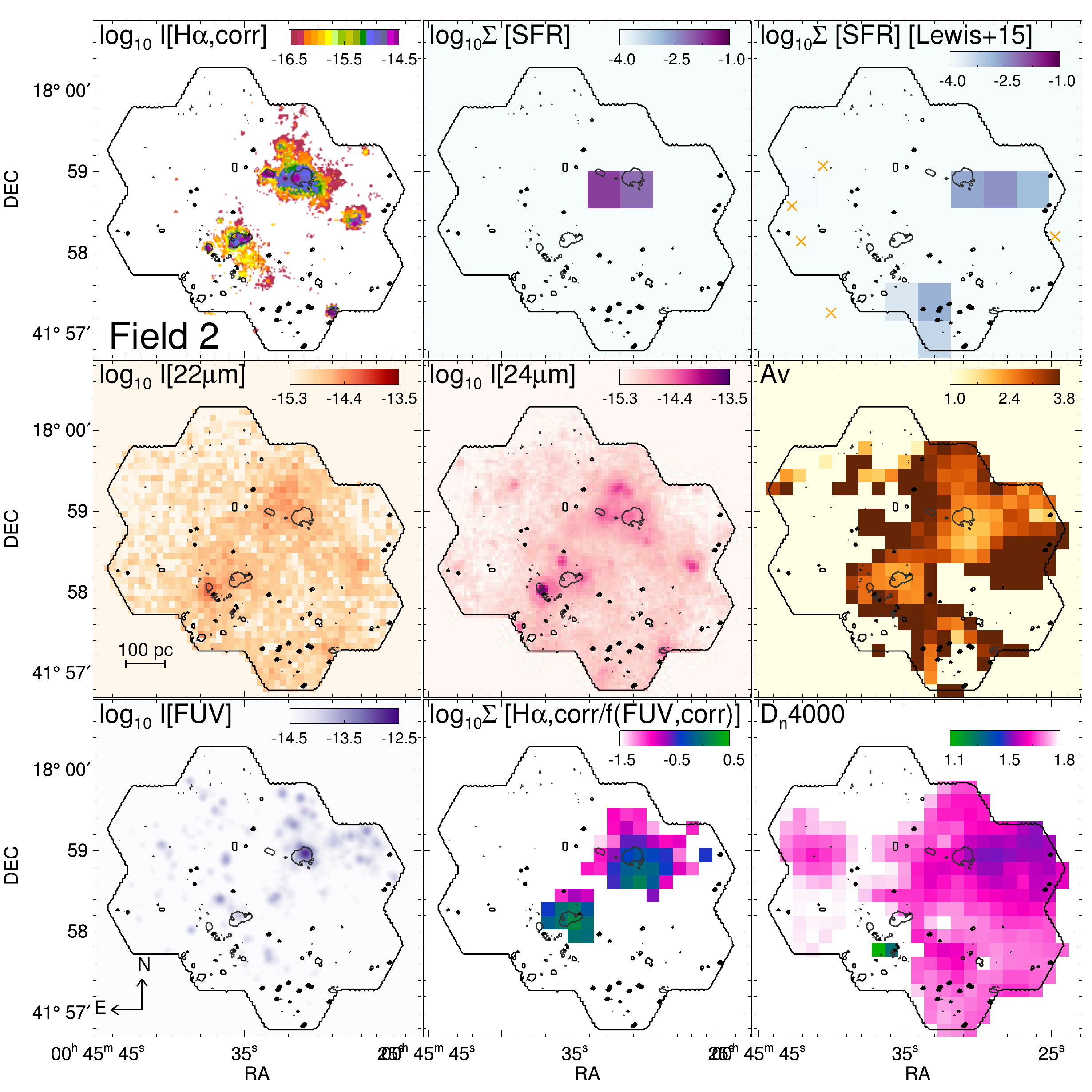}
\caption{ Same as Fig. \ref{fig:SFTmaps_F1}, but for Field 2.  }
\label{fig:SFTmaps_F2}
\end{figure}

\begin{figure}[t!]
\centering
\includegraphics[width=1.0\linewidth]{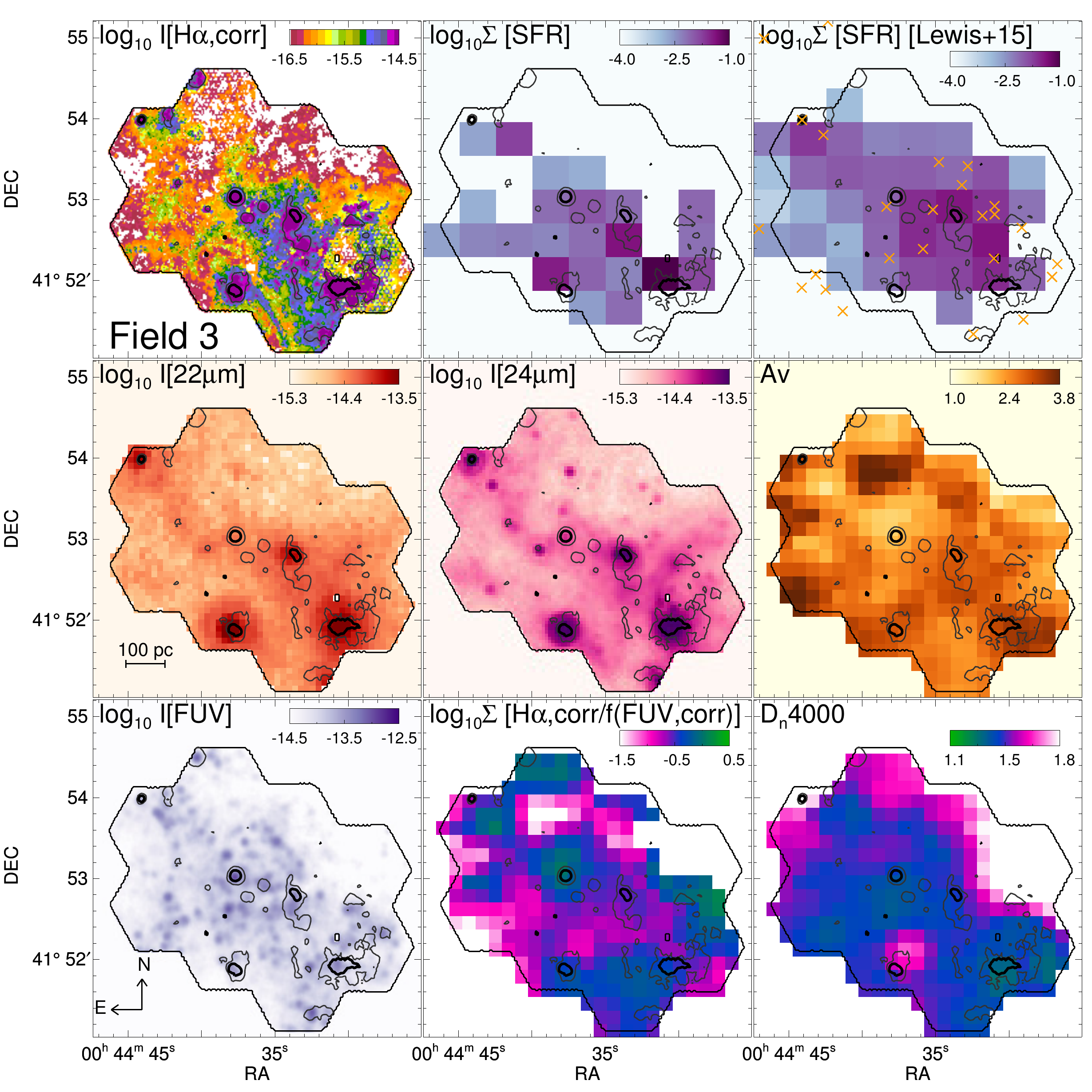}
\caption{ Same as Fig. \ref{fig:SFTmaps_F1}, but for Field 3.  }
\label{fig:SFTmaps_F3}
\end{figure}

\begin{figure}[t!]
\centering
\includegraphics[width=1.0\linewidth]{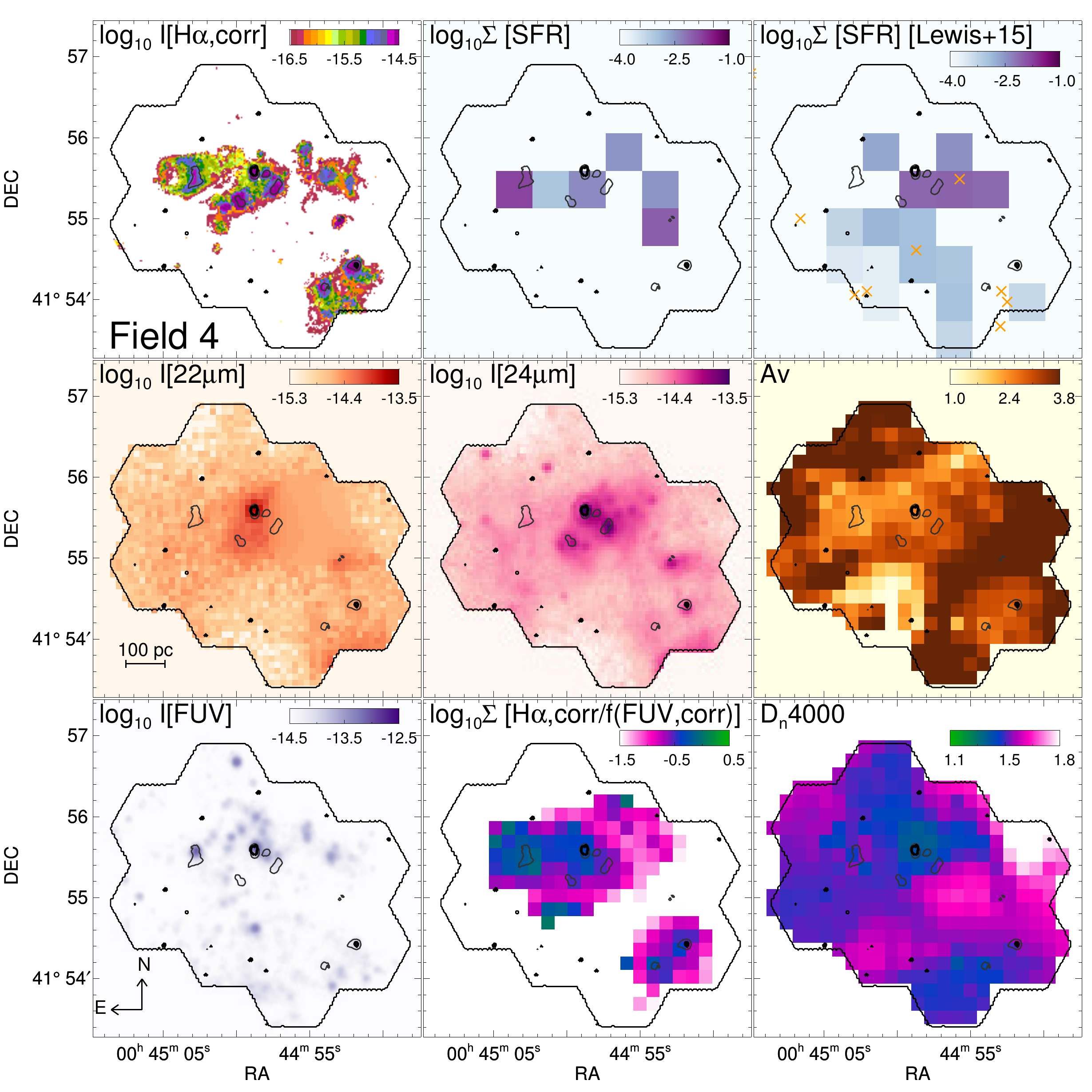}
\caption{ Same as Fig. \ref{fig:SFTmaps_F1}, but for Field 4.  }
\label{fig:SFTmaps_F4}
\end{figure}

\begin{figure}[t!]
\centering
\includegraphics[width=1.0\linewidth]{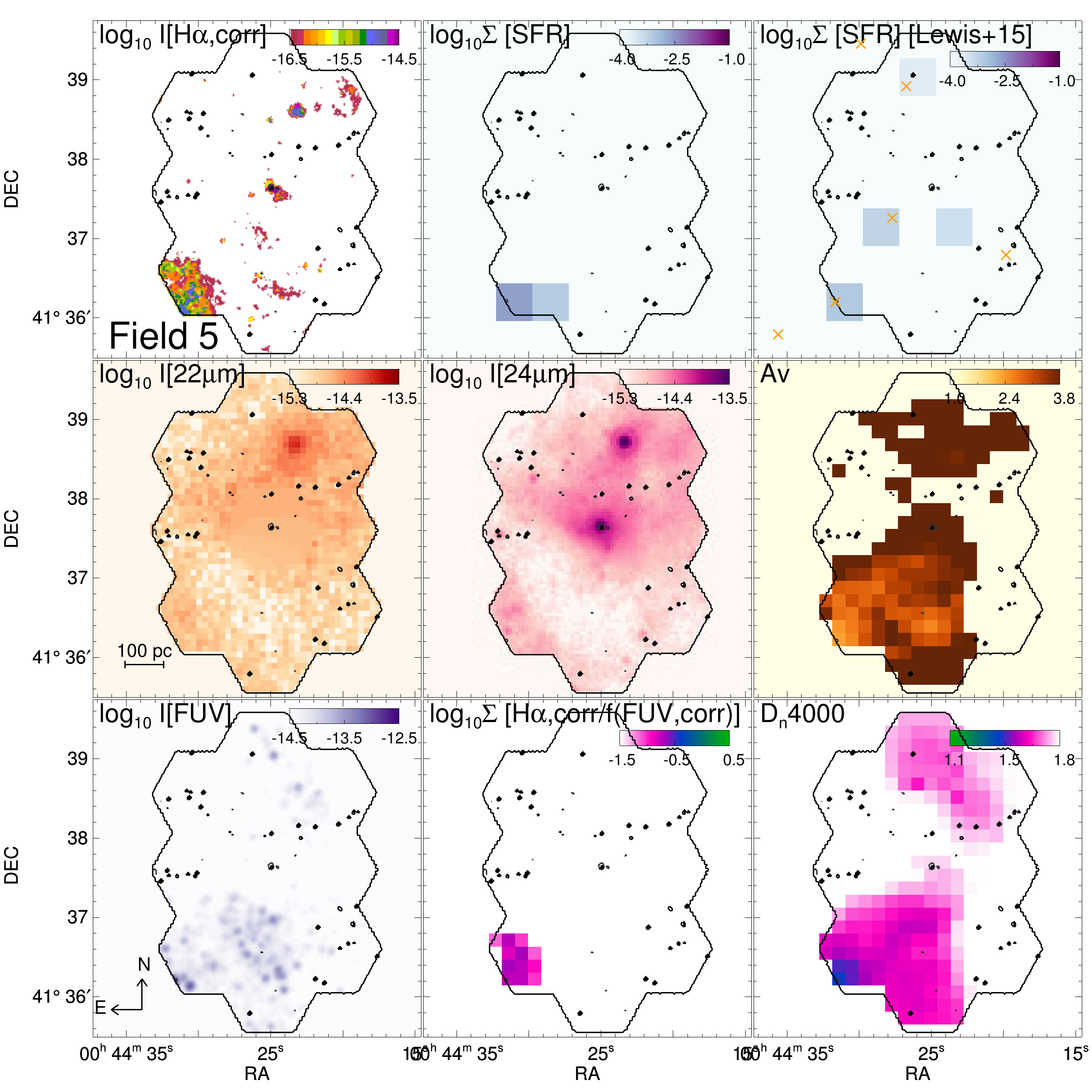}
\caption{ Same as Fig. \ref{fig:SFTmaps_F1}, but for Field 5.  }
\label{fig:SFTmaps_F5}
\end{figure}

In Fig. \ref{fig:SFTmaps_F2}, \ref{fig:SFTmaps_F3}, \ref{fig:SFTmaps_F4} and \ref{fig:SFTmaps_F5}, we show the maps of various tracers and values presented in Fig. \ref{fig:SFTmaps_F1} for Fields 2 to 5. For detailed explanation of the panels, see Fig. \ref{fig:SFTmaps_F1}.  

\clearpage
\newpage

\clearpage
\newpage

\section{SFR prescriptions\label{Sec:Appendix_B_Prescription}}

Here we present the SFR prescriptions derived in this paper, which show results at different spatial scales, and for different tracers (monochromatic and hybrid). The results for H$\alpha$, FUV, 22\,$\mu$m and 24\,$\mu$m  are listed in Tab. \ref{tab:AppB_Tab04}. 

We also added the monochromatic calibration for 12\,$\mu$m, 70\,$\mu$m, 160\,$\mu$m and TIR tracers in Tab. \ref{tab:AppB_Tab05}. TIR values in this work are calculated using Formula 5 in \citet{Dale09}. If we were to use the TIR prescription from \citet{Galametz13}, the TIR values would be $\approx$10\% lower. For TIR, we used $22\,\mu$m instead of $24\,\mu$m values. 
 As seen in Fig. \ref{fig:PACS_TIR}, SFR prescriptions for  70\,$\mu$m and TIR differ from those proposed by \citet{Calzetti10,Calzetti13}, while they are consistent with the 160\,$\mu$m ones from \citet{Calzetti10}. This could
 be related to the fact that the  160\,$\mu$m emission is tracing cold dust, and not the hot dust around star-forming regions.

 There is a disagreement between the log(SFR)-log(IR) correlation derived in this work and prescribed by \citet{Cluver17}. 
 Here,  IR indicates  L(12\,$\mu$m) or  L(22\,$\mu$m). We investigate the shift in the intercept of the  log(SFR)-log(IR) correlation in the case where we  fixed the slope of the power-law fit through the data. 
 Here we assume the slopes from \citet{Cluver17} and list the resulting intercepts in the lower part of Tab. \ref{tab:AppB_Tab05}.  
 We find that the intercepts for 22\,$\mu$m agree with the prescription given by \citet{Cluver17}, while the intercepts for 12\,$\mu$m data have lower values (by $\approx$0.7 dex).

If we apply the fit given by \citet{Cluver17}  to  the M31 data, SFR(12\,$\mu$m) values would be by a factor of 3 higher (0.5 dex) than the SFR(H$\alpha,\rm corr$), regardless of the aperture sizes. 
 \citet{Cluver17} argue that while the  12\,$\mu$m emission contains components of both PAH (Polycyclic aromatic hydrocarbon) emission and continuum emission from hot dust, the PAH emission contributes only $\approx$30\% and therefore the SFR(12\,$\mu$m) prescription is robust. 
However, we suggest that the 12\,$mu$m and 22\,$mu$m tracers are less reliable in M\,31 as the dust composition and distribution probed by our data are different to those observed in other galaxies. 
The  (average) ratios of  $\Sigma$(12\,$\mu$m)/$\Sigma$(22\,$\mu$m) are $ \approx3\pm1 $ in M\,31, while \citet{Cluver17} report similar luminosity values for $\nu\rm L_{\nu}$(12\,$\mu$m) and $\nu\rm L_{\nu}$(22\,$\mu$m) (see Fig. 4 in their paper). 
Only in the very bright HII regions in M\,31 we find a low ratio (of $\approx$1) with an increase towards the outer edges of the regions. 
Interestingly, the 12\,$\mu$m/22\,$\mu$m ratios are weakly anti-correlated with the 70\,$\mu$m/160\,$\mu$m ratios in our fields. 
A similar behavior is seen in nearby galaxies where \citet{Soifer91}  and \citet{Sanders03} found a strong anti-correlation of 12\,$\mu$m/25\,$\mu$m ratios with 60\,$mu$m/100\,$\mu$m ratios and with IR surface brightness which the authors explain by the destruction of small grains in bright mid-IR emission regions.  

\clearpage
\newpage

\begin{table*}[t!]
\centering
\caption{ SFR prescriptions derived for various observed tracers (left column) at different spatial resolutions.  Prescriptions for pixel-by-pixel comparison of star formation tracers maps (middle column), and different aperture sizes (right column) are given. 
For all reported prescriptions diffuse emission has \textit{not} been removed (see Sec. \ref{Subsec:ScalesAndDiff} for removal of diffuse emission). 
We show a mean values of the  calibration factors, and the standard deviation of their  scatter. 
Star formation tracers are given in surface brightness  (with units of $\rm erg\,s^{-1}\,kpc^{-2}$).
Mid-IR (FUV) fluxes are in units of $\rm \nu\,F_{\nu} (\lambda\,F_{\lambda})$, where $\nu$ ($\lambda$) is the effective frequency (wavelength), and $\rm F_{\nu} (F_{\lambda})$ the corresponding flux density.  
H$\alpha,\rm corr$-SFR  and  FUV,corr-SFR  conversion factors a$_{\rm H\alpha,\rm corr}$ and b$_{\rm FUV,corr}$ are stated in the formula \ref{eq:Eq01} and \ref{eq:Eq02}. }
\begin{tabular}{ccc}
\\
SF Tracer &  pixel-by-pixel (native IR resolution) & apertures (R$ \approx $50\,pc)\\ 
\hline 
\\
$ \Sigma_{\rm SFR}$(24$ \mu $m)= a$\times\Sigma_{\rm 24\,\mu \rm m}$ & a$=1.8\pm0.9 \times10^{-42} $ &  a$=1.4\pm0.8 \times10^{-42} $\\

$ \Sigma_{\rm SFR}$(22$ \mu $m)= a$\times\Sigma_{\rm 22\,\mu \rm m}$    & a$=1.5\pm0.9 \times10^{-42} $ &  a$=1.2\pm0.6 \times10^{-42} $\\

$ \Sigma_{\rm SFR}$(H$\alpha$+24$ \mu $m)=a$_{\rm H\alpha,\rm corr}\cdot(\Sigma_{\rm H\alpha}$+a$_{24}\times\Sigma_{\rm 24\,\mu \rm m})$  & a$_{24}=0.24\pm0.14$ &            a$_{24}=0.19\pm0.14$  \\

$ \Sigma_{\rm SFR}$(H$\alpha$+22$ \mu $m)=a$_{\rm H\alpha,\rm corr}\cdot(\Sigma_{\rm H\alpha}$+a$_{22}\times\Sigma_{\rm 22\,\mu \rm m})$ & a$_{22}=0.21\pm0.12$ &            a$_{22}=0.17\pm0.1$  \\

$ \Sigma_{\rm SFR}$(FUV+24$ \mu $m)=b$_{\rm FUV,corr}\cdot(\Sigma_{\rm FUV}$+b$_{24}\times\Sigma_{\rm 24\,\mu \rm m})$ & b$_{24}=33\pm21$ &            b$_{24}=27\pm19$  \\

$ \Sigma_{\rm SFR}$(FUV+22$ \mu $m)=b$_{\rm FUV,corr}\cdot(\Sigma_{\rm FUV}$+b$_{22}\times\Sigma_{\rm 22\,\mu \rm m})$ & b$_{22}=28\pm17$ &            b$_{22}=24\pm13$  \\
\\
SF Tracer &  pixel-by-pixel (25'' resolution) & apertures (R$ \approx $100\,pc)\\ 
\hline 
$ \Sigma_{\rm SFR}$(24$ \mu $m)= a$\times\Sigma_{\rm 24\,\mu \rm m}$ & a$=1.3\pm0.5 \times10^{-42} $ &  a$=1.3\pm0.5 \times10^{-42} $\\

$ \Sigma_{\rm SFR}$(22$ \mu $m)= a$\times\Sigma_{\rm 22\,\mu \rm m}$    & a$=1.2\pm0.6 \times10^{-42} $ &  a$=1.1\pm0.5 \times10^{-42} $\\

$ \Sigma_{\rm SFR}$(H$\alpha$+24$ \mu $m)=a$_{\rm H\alpha,\rm corr}\cdot(\Sigma_{\rm H\alpha}$+a$_{24}\times\Sigma_{\rm 24\,\mu \rm m})$  & a$_{24}=0.18\pm0.08$ &            a$_{24}=0.22\pm0.08$  \\

$ \Sigma_{\rm SFR}$(H$\alpha$+22$ \mu $m)=a$_{\rm H\alpha,\rm corr}\cdot(\Sigma_{\rm H\alpha}$+a$_{22}\times\Sigma_{\rm 22\,\mu \rm m})$ & a$_{22}=0.180.1$ &            a$_{22}=0.17\pm0.09$  \\

$ \Sigma_{\rm SFR}$(FUV+24$ \mu $m)=b$_{\rm FUV,corr}\cdot(\Sigma_{\rm FUV}$+b$_{24}\times\Sigma_{\rm 24\,\mu \rm m})$ & b$_{24}=33\pm21$ &            b$_{24}=25\pm12$  \\

$ \Sigma_{\rm SFR}$(FUV+22$ \mu $m)=b$_{\rm FUV,corr}\cdot(\Sigma_{\rm FUV}$+b$_{22}\times\Sigma_{\rm 22\,\mu \rm m})$ & b$_{22}=24\pm14$ &            b$_{22}=21\pm11$  \\
\\
SF Tracer &  pixel-by-pixel (65'' resolution) & apertures (R$ \approx $200\,pc)\\ 
\hline 
\\
$ \Sigma_{\rm SFR}$(24$ \mu $m)= a$\times\Sigma_{\rm 24\,\mu \rm m}$ & a$=1.2\pm0.3 \times10^{-42} $ &  a$=1.1\pm0.3 \times10^{-42} $\\

$ \Sigma_{\rm SFR}$(22$ \mu $m)= a$\times\Sigma_{\rm 22\,\mu \rm m}$    & a$=1.1\pm0.3 \times10^{-42} $ &  a$=1\pm0.3 \times10^{-42} $\\

$ \Sigma_{\rm SFR}$(H$\alpha$+24$ \mu $m)=a$_{\rm H\alpha,\rm corr}\cdot(\Sigma_{\rm H\alpha}$+a$_{24}\times\Sigma_{\rm 24\,\mu \rm m})$  & a$_{24}=0.19\pm0.06$ &            a$_{24}=0.16\pm0.05$  \\

$ \Sigma_{\rm SFR}$(H$\alpha$+22$ \mu $m)=a$_{\rm H\alpha,\rm corr}\cdot(\Sigma_{\rm H\alpha}$+a$_{22}\times\Sigma_{\rm 22\,\mu \rm m})$ & a$_{22}=0.16\pm0.05$ &            a$_{22}=0.13\pm0.05$  \\

$ \Sigma_{\rm SFR}$(FUV+24$ \mu $m)=b$_{\rm FUV,corr}\cdot(\Sigma_{\rm FUV}$+b$_{24}\times\Sigma_{\rm 24\,\mu \rm m})$ & b$_{24}=25\pm8$ &            b$_{24}=23\pm6$  \\

$ \Sigma_{\rm SFR}$(FUV+22$ \mu $m)=b$_{\rm FUV,corr}\cdot(\Sigma_{\rm FUV}$+b$_{22}\times\Sigma_{\rm 22\,\mu \rm m})$ & b$_{22}=22\pm7$ &            b$_{22}=20\pm6$  \\
\\
SF Tracer &  Integrated field  &\\
\hline 
$ \Sigma_{\rm SFR}$(H$\alpha$+24$ \mu $m)=a$_{\rm H\alpha,\rm corr}\cdot(\Sigma_{\rm H\alpha}$+a$_{24}\times\Sigma_{\rm 24\,\mu \rm m})$  & a$_{24}=0.17\pm0.04$ &           \\

$ \Sigma_{\rm SFR}$(H$\alpha$+22$ \mu $m)=a$_{\rm H\alpha,\rm corr}\cdot(\Sigma_{\rm H\alpha}$+a$_{22}\times\Sigma_{\rm 22\,\mu \rm m})$ & a$_{22}=0.15\pm0.04$ &          \\

$ \Sigma_{\rm SFR}$(FUV+24$ \mu $m)=b$_{\rm FUV,corr}\cdot(\Sigma_{\rm FUV}$+b$_{24}\times\Sigma_{\rm 24\,\mu \rm m})$ & b$_{24}=20\pm6$ &         \\

$ \Sigma_{\rm SFR}$(FUV+22$ \mu $m)=b$_{\rm FUV,corr}\cdot(\Sigma_{\rm FUV}$+b$_{22}\times\Sigma_{\rm 22\,\mu \rm m})$ & b$_{22}=19\pm5$ &            \\
\\
\end{tabular}  \\
\label{tab:AppB_Tab04}
\end{table*}

 \begin{table*}[h!]
\centering
\caption{ SFR prescriptions derived from for 12\,$\mu$m, 22\,$\mu$m, 70\,$\mu$m, 160\,$\mu$m and TIR tracers (left column) for different aperture sizes (radii of 50\,pc, 100\,pc and 200\,pc). For values provided in the
lower part of the table, the slope of the log(SFR)-log(IR) relations has been fixed to the value calibrated by \citet{Cluver17}. In that case, \citet{Cluver17} prescribe the intercept to be -8 (-7.8) for the 22\,$\mu$m (12\,$\mu$m) tracer. 
For all provided prescriptions diffuse emission has \textit{not} been removed (see Sec. \ref{Subsec:ScalesAndDiff} for removal of diffuse emission).
In the square brackets, we report the scatter of the calibration factors. 
Star formation tracers are given in surface brightness  (with units of $\rm erg\,s^{-1}\,kpc^{-2}$).
IR fluxes are in units of $\rm \nu\,F_{\nu} (\lambda\,F_{\lambda})$, where $\nu$ ($\lambda$) is the effective frequency (wavelength), and $\rm F_{\nu} (F_{\lambda})$ the corresponding flux density.  }
\begin{tabular}{cccc}
\\
SF Tracer &  R=50\,pc &  R=100\,pc  &   R=200\,pc   \\ 
\hline 

$ log_{10}$[SFR(22$ \mu $m)]= a$\times log_{10}[\frac{L_{\rm 22\,\mu \rm m}}{L_{\odot}}]$+b & a= 0.3$\pm$0.1, b= -5.3 & a= 0.5$\pm$0.2, b= -6.3  & a= 1.1$\pm$0.2, b= -9.1 \\

$ log_{10}$[SFR(12$ \mu $m)]= a$\times log_{10}[\frac{L_{\rm 12\,\mu \rm m}}{L_{\odot}}]$+b & a= 0.9$\pm$0.1, b= -8.3 & a= 0.9$\pm$0.2, b= -8.4  & a= 1.2$\pm$0.5, b= -10.1 \\

$ \Sigma_{\rm SFR}$(12$ \mu $m)= a$\times\Sigma_{\rm 12\,\mu \rm m}$ & a=$4.7\pm2.8\times10^{-43} $ & a=$4.1\pm1.8\times10^{-43} $ &  a=$3.6\pm1.7\times10^{-43} $ \\

$ \Sigma_{\rm SFR}$(70$ \mu $m)= a$\times\Sigma_{\rm 70\,\mu \rm m}$ & a=$4.8\pm4.3\times10^{-43} $ & a=$3.7\pm1.9times10^{-43} $ & a=$3.5\pm1.9\times10^{-43} $  \\

$ \Sigma_{\rm SFR}$(160$ \mu $m)= a$\times\Sigma_{\rm 160\,\mu \rm m}$ & a=$1.8\pm0.9\times10^{-43} $  & a=$1.6\pm0.7\times10^{-43} $ & a=$1.4\pm0.6\times10^{-43} $  \\

$ \Sigma_{\rm SFR}$(TIR)= a$\times\Sigma_{\rm TIR\,\mu \rm m}$ & a=$9.3\pm4.8\times10^{-44} $ & a=$8\pm3\times10^{-44} $ &  a=$7\pm3\times10^{-44} $ \\

$ \Sigma_{\rm SFR}$=a$_{\rm H\alpha,\rm corr}\cdot(\Sigma_{\rm H\alpha}$+a$_{12}\times\Sigma_{\rm 12\,\mu \rm m})$ & a$_{12}=0.04\pm0.03$ & a$_{12}=0.04\pm0.02$     &  a$_{12}=0.04\pm0.02$  \\

$ \Sigma_{\rm SFR}$=a$_{\rm H\alpha,\rm corr}\cdot(\Sigma_{\rm H\alpha}$+a$_{70}\times\Sigma_{\rm 70\,\mu \rm m})$ & a$_{70}=0.07\pm0.07$ & a$_{70}=0.06\pm0.03$      & a$_{70}=0.06\pm0.07$  \\

$ \Sigma_{\rm SFR}$=a$_{\rm H\alpha,\rm corr}\cdot(\Sigma_{\rm H\alpha}$+a$_{160}\times\Sigma_{\rm 160\,\mu \rm m})$ & a$_{160}=0.03\pm0.02$  &  a$_{160}=0.03\pm0.01$ &  a$_{160}=0.02\pm0.01$   \\

$ \Sigma_{\rm SFR}$=a$_{\rm H\alpha,\rm corr}\cdot(\Sigma_{\rm H\alpha}$+a$_{\rm TIR}\times\Sigma_{\rm TIR})$ & a$_{\rm TIR}=0.013\pm0.008$  &  a$_{\rm TIR}=0.012\pm0.005$  &  a$_{\rm TIR}=0.011\pm0.004$   \\
\hline 
  Fixed slope (\citealt{Cluver17}): & & & \\

$ log_{10}$[SFR(22$ \mu $m)]= 0.92$\times log_{10}[\frac{L_{\rm 22\,\mu \rm m}}{L_{\odot}}]$+b &  b=$-7.9\pm0.4$ &  b=$-7.9\pm0.3$ &  b=$-7.8\pm0.1$ \\

$ log_{10}$[SFR(12$ \mu $m)]= 0.89$\times log_{10}[\frac{L_{\rm 12\,\mu \rm m}}{L_{\odot}}]$+b & b=$-8.3\pm0.2$ &  b=$-8.3\pm0.2$  &  b=$-8.3\pm0.2$ \\

\\
\end{tabular}  \\
\label{tab:AppB_Tab05}
\end{table*}

\clearpage
\newpage

\section{Calibration factor as a function of various physical quantities\label{Sec:a_vs_others}}

\begin{figure*}[t!]
\centering
\includegraphics[width=1.0\linewidth]{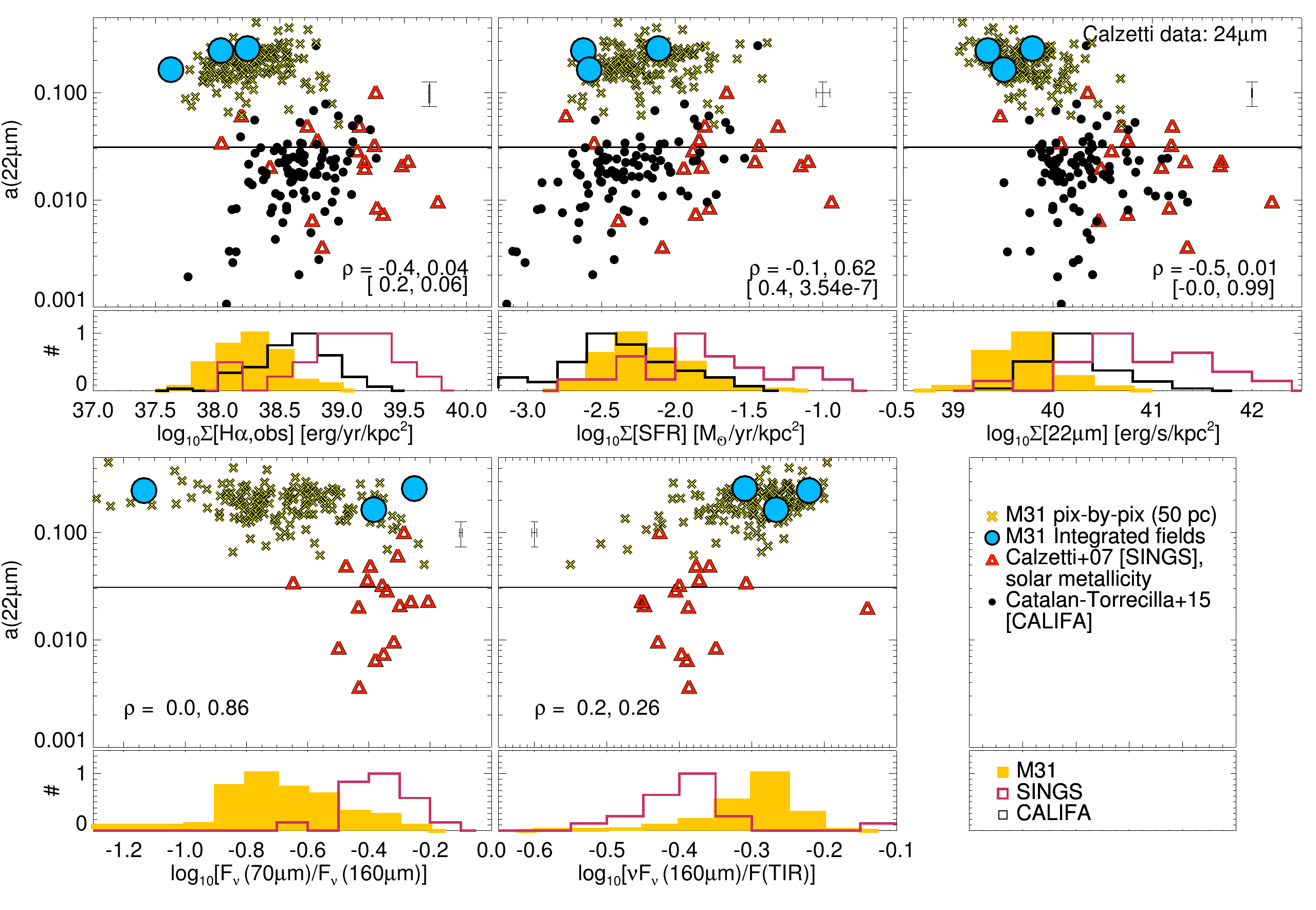}
\caption{ 
In the diagrams, we show  a$_{\rm 22}$, estimated individually for each  data point, as a function of   observed $\Sigma(\rm H\alpha)$ (top left), $\Sigma(\rm SFR)$ (top middle), $\Sigma(\rm 22\,\mu\rm m)$ (top right), $70\,\mu\rm m/160\,\mu\rm m$ ratio (indicating the dust temperature; bottom left), and  $160\,\mu\rm m/TIR$ ratio (tracing contribution from cold dust emission to TIR; bottom middle).  
Data points in upper panels come from:  M31 pixel-by-pixels data (spatially independent pixels with 50\,pc length; yellow X symbols), integrated fields (blue circles), the data of the central square regions in SINGS galaxies (spatial base lengths of 140\,pc-4\,kpc) from C07 with  metallicities comparable with M31 (red triangles; C07 and  \citealt{Dale17}), and the aperture data from the CALIFA galaxies from \citet{CatalanTorrecilla15}. 
  A variation in the a$_{\rm 22}$ factor indicates a behavior of the SFR prescription, with a$_{\rm 22}=0.031$ from the C07 prescription indicated with the black solid line.   
  The histograms bellow diagrams show the distribution of M31 data (filled yellow histograms), the SINGS galaxies data (red empty histograms) and CALIFA galaxies (black empty histograms)  as a function of corresponding quantities on the x-axis in upper diagrams.  
   The estimated uncertainties are shown, as the Spearman's  correlation coefficient ($ \rho $, left number) and  the significance of its deviation from zero (right number)  for the integrated M31 fields and SINGS  data (upper numbers), and for the integrated M31 fields, SINGS and CALIFA data (numbers in brackets).
    In these diagrams and histograms, we see a slight trend in a$_{\rm 22}$ with  mid-IR emission, the dust temperature, and with $160\,\mu\rm m/TIR$ ratio.  For details, see the text.   
    }
\label{fig:a22_vs_others}
\end{figure*}

 Fig. \ref{fig:a22_vs_others} presents a$_{22}$, estimated for each individual data point,  as a function of physical quantities: observed $\Sigma(\rm H\alpha)$, $\Sigma(\rm SFR)$,  $\Sigma(\rm 22\,\mu\rm m)$, $70\,\mu\rm m/160\,\mu\rm m$ ratio (tracing dust temperature), and  $160\,\mu\rm m/TIR$ ratio(tracing cold gas contribution to TIR).
 A histogram of the data distribution for each corresponding physical quantity is plotted below each panel.
 We estimated the Spearman's  correlation coefficient ($ \rho $) and the corresponding  significance of its deviation from zero for the integrated M31 fields and SINGS  data (upper numbers), and for the integrated M31 fields, SINGS and CALIFA data (numbers in brackets). 
 The $ \rho $ factor that is estimated by combining the M31's integrated fields and the SINGS data is more reliable as those data probe similar spatial scales, unlike the resulting factor that includes CALIFA data as well as these data may  probe entire galaxies.

 The figure shows our M31 data points, SINGS galaxies with metallicities similar to M31 (used by C07 for their SFR calibrations), and CALIFA survey galaxies (used by \citealt{CatalanTorrecilla15}). 
    For M31, we plot integrated fields and the  pixel-by-pixel comparison of the maps at $\rm 22\,\mu\rm m$ resolution. 
 Here, we binned the pixels to a size of 50\,pc for spatially independent measurements.  
 The data provided by C07 and shown in Fig. \ref{fig:a22_vs_others} are the central square  regions in SINGS galaxies, with a spatial base length  between 140\,pc and 4\,kpc.  
 The mid-IR surface brightness for the C07 data points would yield  slightly higher values if diffuse emission is not removed.
 The PACS $70\,\mu$m and 160\,$\mu$m measurements of C07 are derived at galactic scales and taken from \citet{Dale17}. 
  The data from  the CALIFA survey are from apertures (with 36'' radius) covering entire or most of galaxies.

 The data and the histograms show that our M31's fields exhibit  an  order of magnitude lower $\Sigma(\rm H\alpha, obs)$ and $\Sigma(\rm IR)$ and a slightly lower $\Sigma(\rm SFR)$  compared to the C07 data. 
We notice a slight trend (anti-correlation) between a$_{\rm IR}$ and $\Sigma(\rm IR)$, $\Sigma(\rm H\alpha, obs)$, and a slight correlation with the $160\,\mu\rm m/TIR$ ratio. 
These trends are also seen in the $\rho$ factors for M31's integrated fields and the SINGS galaxies.  
Although, there is slight anti-correlation between a$_{\rm IR}$ and $70\,\mu\rm m/160\,\mu\rm m$ ratio (probing dust temperature) when we compare pixel-by-pixel data in M31 and the SINGS data, this anti-correlation breaks down for the integrated fields in M31. 
We explain this as an effect of HII region emission dominating the emission within the integrated  fields.  
However, we only have a data points from only 3 fields, with one field showing much lower dust temperature than others, which is a small number to draw a strong conclusions about the dust temperature in the M31's fields. 
The histograms also indicate that  the pixel-by-pixel M31 data have lower $70\,\mu\rm m/160\,\mu\rm m$ ratios than the SINGS galaxies.

\end{document}